\documentclass[usenatbib]{mn2e}
\usepackage{times}
\usepackage{natbib}
\usepackage{epsfig}
\usepackage{bm}

\newcommand{\bdv}[1]{{\bf #1}}

\newcommand{\ds}{\delta_s}     
\newcommand{\MX}{\xi^s}        
\newcommand{\OM}{\Omega_m}
\newcommand{\rms}{\sigma_8}

\newcommand{\RA}{\rightarrow}
\newcommand{\beeq}{\begin{equation}}
\newcommand{\VV}{\mathcal{V}}  
\newcommand{\dm}{\delta_m}     
\newcommand{\eneq}{\end{equation}}
\newcommand{\HH}{\mathcal{H}}  
\newcommand{\dr}{\delta_g}     
\newcommand{\bng}{\bar n_g}    
\newcommand{\kvec}{\bdv{k}}
\newcommand{\svec}{\bdv{s}}
\newcommand{\zvec}{\bdv{z}}
\newcommand{\Xang}{\hat\bdv{x}}
\newcommand{\Kang}{\hat\bdv{k}}
\newcommand{\PZ}{P_s}          
\newcommand{\PM}{P_m}          
\newcommand{\LL}{\mathcal{P}}  
\newcommand{\MP}{P^s}          
\newcommand{\bear}{\begin{eqnarray}}
\newcommand{\enar}{\end{eqnarray}}
\newcommand{\XZ}{\xi_s}        
\newcommand{\Vang}{\bdv{\hat n}}
\newcommand{\up}[1]{{\rm #1}}
\newcommand{\oo}{\Theta}       

\newcommand{\mpc}{\up{Mpc}}
\newcommand{\gpc}{\up{Gpc}}
\newcommand{\hmpc}{{h^{-1}\mpc}}
\newcommand{\hgpc}{{h^{-1}\gpc}}
\newcommand{\af}{\alpha_\up{full}} 
\newcommand{\fsky}{f_\up{sky}} 
\newcommand{\PP}{P_z}          
\newcommand{\pdf}{P}           
\newcommand{\kmin}{k_\up{min}}
\newcommand{\TMP}{\tilde\MP_0} 
\newcommand{\kms}{{\rm km\, s}^{-1}}
\newcommand{\hmpci}{{h\mpc^{-1}}}

\begin{document}

\title{Wide angle effects in future galaxy surveys}
\author[Yoo \& Seljak]{Jaiyul Yoo$^{1,2}$\thanks{email: 
jyoo@physik.uzh.ch,~~ jyoo@lbl.gov}
and Uro{\v s} Seljak$^{1,2,3,4}$\\
$^1$Institute for Theoretical Physics, University of Z\"urich,
CH-8057 Z\"urich, Switzerland\\
$^2$Lawrence Berkeley National Laboratory, University of 
California, Berkeley, CA 94720, USA\\
$^3$Physics Department and Astronomy Department,
University of California, Berkeley, CA 94720, USA\\
$^4$Institute for the Early Universe, Ewha Womans University,
120-750 Seoul, South Korea
}

\maketitle

\begin{abstract}
Current and future galaxy surveys cover a large fraction of the entire sky
with a significant redshift range, and the recent theoretical development
shows that general relativistic effects are present in galaxy clustering
on very large scales. This trend has renewed interest in the
wide angle effect in galaxy clustering measurements, in which the 
distant-observer approximation is often adopted. 
Using the full wide-angle formula for computing the redshift-space
correlation function, we show that 
compared to the sample variance, the deviation in the redshift-space 
correlation function from the simple
Kaiser formula with the distant-observer approximation 
is negligible in galaxy surveys such as the SDSS, 
Euclid and the BigBOSS, if the theoretical prediction 
from the Kaiser formula is properly averaged over the survey volume.
We also find corrections to the wide-angle formula
and clarify the confusion in literature between the wide angle effect and the 
velocity contribution in galaxy clustering. 
However, when the FKP method is applied, substantial deviations
can be present in the power spectrum analysis in future surveys, 
due to the non-uniform distribution of galaxy pairs.
\end{abstract}


\section{Introduction}
\label{sec:intro}

In the past two decades, rapid experimental developments in large-scale
galaxy surveys have revolutionised our understanding of the Universe
such as the Sloan Digital Sky Survey (SDSS; \citealt{YOADET00}),
the Two degree Field Galaxy Redshift Survey (2dFGRS; \citealt{CODAET01}),
and the WiggleZ Dark Energy Survey \citep{DRJUET10}.
In particular, 
matter fluctuations on large scales remain in the linear regime, where
it is simple to relate the cosmological measurements
to the governing cosmological parameters. Since two-point statistics
provides a complete description of the Gaussian random field, a great deal
of effort has been devoted to measuring the correlation function in 
configuration space and the power spectrum in Fourier space (e.g., 
\citealt{TEBLET04,EIZEET05,COPEET05}). The current state-of-the-art 
measurements are \cite{REPEET10} power spectrum 
analysis of the SDSS luminous red galaxy (LRG) samples,
\cite{BLDAET11} correlation function analysis of the WiggleZ survey,
and \cite{SASCET12} correlation function analysis of the Baryonic
Oscillation Spectroscopic Survey (BOSS; \citealt{SCBLET07}).
Moreover, future galaxy surveys such as 
Euclid\footnote{http://sci.esa.int/euclid} and the 
BigBOSS\footnote{http://bigboss.lbl.gov} are planned to measure galaxies
at higher redshift with larger sky coverage. With enormous statistical 
power in these future surveys, theoretical predictions of the redshift-space
correlation function and the power spectrum need to be further refined
to take full advantage of high-precision measurements.

Motivated by these recent developments,
the relativistic description of galaxy clustering has been developed to meet
the high accuracy of theoretical predictions demanded by these surveys
(\citealt*{YOFIZA09}; \citealt{YOO10}). 
While measurements of
galaxy clustering are based on observed quantities such as the observed 
redshift and the galaxy position on the sky, its theoretical prediction is
based on unobservable quantities such as the real-space redshift and
unlensed galaxy position. The relativistic formula for the observed galaxy 
fluctuation is constructed by using the observed quantities, 
providing a complete description of all the effects in galaxy clustering
to the linear order in perturbations 
\citep{YOFIZA09,YOO10,CHLE11,BODU11,JESCHI12}. 
The relation between the full Kaiser formula and the relativistic formula
was clarified in \cite{YOHAET12} with a careful examination of 
gauge issues and the Newtonian correspondence. These theoretical developments
reveal that the relativistic effect is present in galaxy clustering,
and the relativistic effect itself provides new opportunities 
to test general relativity on horizon scales in future galaxy surveys
\citep{YOHAET12}.

This recent trend has renewed interest in the wide angle effect in galaxy
clustering measurements on large scales. The standard method for computing
the redshift-space correlation function and its power spectrum on large
scales is based on linear theory and is described by the Kaiser formula
\citep{KAISE87}. On large scales, where linear theory is applicable, 
the Kaiser formula provides a simple and physically transparent relation 
between the real-space $\delta_g$ and the redshift-space $\ds$
galaxy fluctuations, and
its two-point correlation function \citep{HAMIL92,COFIWE94} or the
power spectrum \citep{KAISE87} can be readily computed. 
This simple relation
is made possible by adopting the distant-observer approximation, in which
the observer is far away from galaxies in observation and hence the position
angles of those galaxies are virtually identical on the sky.

Therefore, it is natural to expect this assumption to break down in 
wide-angle galaxy surveys, demanding a formalism for computing the 
redshift-space correlation function without the distant-observer 
approximation. When the distant-observer approximation is dropped,
there exist three different angles, two lines-of-sight directions to the
galaxy pair and the pair separation direction, and the resulting correlation
function in redshift space is an infinite sum of plane waves with different
angular momenta (see also, \citealt{FISCLA94,FILAET95,HETA95,HACU96}
for the spherical power spectrum approach to redshift-space distortion).
Exploiting the fact that the geometry in question is confined to a plane,
\citet*{SZMALA98} first developed a simple expression for computing the 
redshift-space correlation function by using the bi-polar spherical harmonics.
The wide angle formalism is further completed in \citet{SZAPU04} by noting
that the correlation
function configuration specified by three angles can be expanded in terms
of tri-polar spherical harmonics and the total angular momentum of this
expansion must vanish due to rotational invariance. Further extension 
in the wide angle formalism was made by \citet{MATSU00b} to compute the
correlation in a non-flat universe, by \citet{PASZ08} to implement
the full Kaiser formula, by \citet{MODU12} to account for galaxy pairs
at two different redshifts, and by \cite{BEMAET12} to add relativistic 
corrections.

In observational side,
\citet{OKMAET08} analysed the SDSS luminous red galaxy (LRG) sample
to measure the baryonic acoustic oscillation (BAO) scale in two-dimensional 
redshift-space correlation function on large scales and found that
the impact of the wide angle effect on their measurements is small.
\cite{SAPERA12} performed an extensive 
study of systematic errors in interpreting large-scale redshift-space 
measurements in the SDSS. By quantifying the distribution of the opening 
angle as a function of pair separation, they concluded that the wide angle 
effect is negligible in the SDSS.
\citet{BEBLET11} used the 6dF Galaxy Survey 
for their BAO measurements by using the angle-averaged monopole 
correlation function. They concluded that the wide angle effect on the 
monopole correlation function
measurements is minor: $\Delta\xi^s_0\simeq10^{-4}$ at the BAO scale, much
smaller than the measurement uncertainties.

However, \citet{RABEET12} argue that the wide angle effect in galaxy
clustering measurements is potentially degenerate with the signature of
modified gravity models and it should be considered interpreting measurements
 in future galaxy surveys. 
For galaxy pairs that are widely separated in angle,
the redshift-space correlation function is sufficiently different from that
obtained by using the simple Kaiser formula with the distant-observer
approximation, and the deviation in the correlation function measurements
might be misinterpreted as the breakdown of general relativity.
However, at large opening angles, where the wide angle effect is largest,
there are few galaxy pairs at a typical pair separation, 
and the measurement uncertainties are 
larger. Therefore, it is important to quantify the measurement uncertainties
associated with galaxy pairs at large opening angles, and it is described
by the probability distribution of the triangular shapes of galaxy
pairs in each survey.

Here we perform a systematic study of the wide angle effect in galaxy
clustering measurements in galaxy surveys such as the SDSS, Euclid
and the BigBOSS. Our results on the wide angle effect in the SDSS agree
with the previous work in \cite{SAPERA12}. On the other hand, there is
{\it no} systematic study on the wide angle effect in future galaxy surveys.
The lack of study in this direction is partially due to the
difficulty in predicting measurement uncertainties in the correlation
function, while various ways exist to achieve this goal
when measurements are already made such as in the SDSS
(see \citealt{SAPERA12} for details).
In contrast, it is easier to predict measurement
uncertainties in the power spectrum, as each Fourier mode is independent on
large scales. Therefore, our strategy is as follows.
We first quantify the probability distribution of the triangular
shapes formed by the observer and galaxy pairs that fit in the survey regions.
The redshift-space correlation function can then be obtained by averaging
each correlation function over all triangular configurations. Second,
we compute the redshift-space power spectrum by Fourier transforming 
the the resulting redshift-space correlation function, and we quantify the
systematic errors in theoretical predictions based on
the simple Kaiser formula with the distant-observer approximation
by computing the covariance matrix of the redshift-space power spectrum.

Our eventual conclusion is that galaxies in surveys are sufficiently
far away from the observer, and the distant-observer approximation is highly
accurate. However, one has to be careful to avoid significant systematic
errors when comparing theoretical predictions with measurements,
because the Kaiser formula with the distant-observer approximation needs to
be evaluated at a certain redshift and the deviation from the
uniform distribution of cosine angle~$\mu$ is substantial on large pair
separations. Once these issues are properly considered in estimating the
redshift-space correlation function, the systematic errors from the use of
the Kaiser formula are negligible in the SDSS and are completely irrelevant
in future survey.
Furthermore, 
since the power spectrum analysis in practice is performed in a
slightly different way, we investigate the systematic errors associated with
the present power spectrum analysis, which turn out to be larger than
the wide angle effect.

The outline of the paper is as follows. In Section~\ref{sec:formalism}
the redshift-space distortion formalism is discussed. We first introduce the 
Kaiser formula for the redshift-space galaxy fluctuation and its power
spectrum, and we present the multipole expansion and the covariance matrix.
Second, we briefly review the wide angle formalism for computing the
redshift-space correlation function and clarify the ``wide angle effects,''
i.e., the deviation from the simple Kaiser formula with the distant-observer
approximation. We then make connection to the full 
relativistic formula for the observed galaxy fluctuation and identify
missing velocity corrections in the Kaiser formula and the wide angle
formalism. In Section~\ref{sec:result} we present our main results on 
the systematic errors that may occur in measuring the redshift-space 
correlation function and the power spectrum 
in the SDSS, Euclid, and the BigBOSS
by using the simple Kaiser formula with the distant-observer approximation.
Finally, we conclude in Section~\ref{sec:discussion} with a discussion
of our findings. For illustrations, we adopt a flat $\Lambda$CDM cosmology 
with $\OM=0.27$, $h=0.703$, $n_s=0.966$, and $\rms=0.809$. Throughout the
paper, we only use linear theory, valid on large scales.

\section{Redshift-space distortion}
\label{sec:formalism}
Here we briefly discuss the formalism for redshift-space distortion 
necessary for computing the correlation function and the power spectrum
in Section~\ref{ssec:formalism} and for computing their multipole expansion
and covariance matrix in Section~\ref{ssec:multipole}. We then 
review the extension of the Kaiser formula to all sky and its deviation
from the distant-observer approximation in Section~\ref{ssec:wide}
and make a connection to the full relativistic formula for the observed
galaxy fluctuation in Section~\ref{ssec:gr}.

\subsection{Formalism}
\label{ssec:formalism}
The observed distance~$s$ of a galaxy 
in redshift space is based on the observed 
redshift~$z$ and differs from the real-space 
distance~$r$ due to the line-of-sight peculiar velocity~$V$ as
\beeq
s=r+\VV=r+f{\partial\over\partial r}\nabla^{-2}\dm~,
\label{eq:rr}
\eneq
where the comoving line-of-sight displacement is $\VV=V/\HH$, the logarithmic
growth rate is~$f=d\ln D/d\ln a$, the growth factor~$D(z)$ is normalized to
unity at present, and the conformal Hubble parameter is
$\HH=aH$.
Using the conservation of the total number of the observed galaxies
in a small volume,
$n_g(s)d^3s=n_g(r)d^3r$, the observed galaxy fluctuation~$\ds$ 
in redshift space is related to the real-space fluctuation~$\dr$ as
\beeq
1+\delta_s={n_g(r)\over n_g(s)}\left|{d^3s\over d^3r}\right|^{-1}
={r^2\bar n_g(r)\over s^2\bar n_g(s)}\left(1+{d\VV\over
dr}\right)^{-1}(1+\dr)~.
\label{eq:conr}
\eneq
This relation is exact but assumes that the redshift-space distortion is purely
radial, ignoring angular displacements.

One can make a progress by expanding equation~(\ref{eq:conr}) to the linear
order in perturbations, and the redshift-space galaxy fluctuation is then
\citep{KAISE87}
\beeq
\ds=\dr-\left({d\over dr}+{\alpha\over r}\right)\VV ~,
\label{eq:fullkaiser0}
\eneq
where the selection function~$\alpha$ is defined in terms of
the (comoving) mean number density $\bng$ of the galaxy sample as
\beeq
\alpha\equiv{d\ln r^2\bng\over d\ln r}=2+{rH\over1+z}
{d\ln\bng\over d\ln(1+z)}~.
\eneq
By adopting the distant-observer approximation ($r\RA\infty$)
and ignoring the velocity contributions, a further simplification can be 
made \citep{KAISE87}:
\beeq
\ds=b~\delta_m-{d\VV\over d r}=
\int{d^3\kvec\over(2\pi)^3}~e^{i\kvec\cdot\svec}~(b+f\mu_k^2)~\dm(\kvec)~,
\label{eq:kaiser0}
\eneq
where we used the linear bias approximation $\dr=b~\dm$ \citep{KAISE84}.
When we consider a fluctuation at one point such as $\ds$, there are no 
ambiguities associated with the line-of-sight direction, and in 
equation~(\ref{eq:kaiser0}) the cosine angle $\mu_k=\hat\svec\cdot\Kang$
between the line-of-sight direction $\hat\svec$ and the wavevector $\kvec$ is 
always well-defined, regardless of the validity of the distant-observer
approximation. It is also noted that 
we ignore the angular displacement due to the gravitational lensing.

Under the distant-observer approximation, all galaxies are far away from the
observer, and their position angles are virtually identical. In this case,
equation~(\ref{eq:kaiser0}) can be used to compute the power spectrum
in redshift space as \citep{KAISE87}
\beeq
\PZ(k,\mu_k)=(b+f\mu_k^2)^2\PM(k)~.
\label{eq:kaiser1}
\eneq
Hereafter, we refer to equations~(\ref{eq:kaiser0}) and~(\ref{eq:kaiser1})
as the Kaiser formulae for the redshift-space galaxy fluctuation~$\ds$
and its power spectrum $\PZ$, but it is noted that 
equation~(\ref{eq:kaiser0}) is the leading order terms $\sim\mathcal{O}(\dm)$
of the full equation~(\ref{eq:fullkaiser0})
based on the distant-observer approximation in linear theory,
and hence equation~(\ref{eq:kaiser1}) $\sim\mathcal{O}(\dm^2)$.
To separate from these ``simple'' Kaiser formulae, we refer to
equation~(\ref{eq:fullkaiser0}) as the full Kaiser formula for the 
redshift-space galaxy fluctuation, but we explicitly spell out to 
distinguish the simple
and the full Kaiser formulae, whenever necessary to avoid confusion.

\subsection{Multipole expansion and covariance matrix}
\label{ssec:multipole}
The simple
Kaiser formula for the redshift-space power spectrum is anisotropic, and
it is often convenient to expand $\PZ(k,\mu_k$) 
in terms of Legendre polynomials
$\LL_l(x)$ as
\beeq
\PZ(k,\mu_k)=\sum_{l=0,2,4}\LL_l(\mu_k)\MP_l(k)~,
\eneq
and the corresponding multipole power spectra are
\beeq
\MP_l(k)={2l+1\over2}\int_{-1}^1 d\mu_k~\LL_l(\mu_k)\PZ(k,\mu_k)~.
\label{eq:multi}
\eneq
With its simple angular structure, 
the simple Kaiser formula in equation~(\ref{eq:kaiser1}) is completely
described by three multipole power spectra
\bear
\label{eq:coef1}
\MP_0(k)&=&\left(b^2+{2fb\over3}+{f^2\over5}\right)\PM(k)~,\\
\MP_2(k)&=&\left({4bf\over3}+{4f^2\over7}\right)\PM(k)~, \\
\label{eq:coef3}
\MP_4(k)&=&{8\over35}f^2\PM(k)~,
\enar
while any deviation from the linearity or the distant-observer approximation
can give rise to higher-order even multipoles ($l>4$)
and deviations of the lowest
multipoles from the above equations.

The correlation function in redshift space is the
Fourier transform of the redshift-space power spectrum $\PZ(k,\mu_k)$.
With the distant-observer approximation the redshift-space
correlation function can
be computed and decomposed in terms of Legendre polynomials as
\beeq
\XZ(s,\mu)=\int{d^3\kvec\over(2\pi)^3}~e^{i\kvec\cdot\svec}~\PZ(k,\mu_k)
=\sum_{l=0,2,4}\LL_l(\mu)~\xi_l^s(s)~,
\label{eq:zcorr}
\eneq
and the multipole correlation functions are related to the multipole power
spectra as \citep{HAMIL92,COFIWE94}
\bear
\label{eq:mul}
\MX_l(s)&=&i^l\int{dk~k^2\over2\pi^2}~\MP_l(k)j_l(ks)~,\\
\label{eq:mulp}
\MP_l(k)&=&4\pi(-i)^l\int~dx~x^2\MX_l(x)j_l(kx)~,
\enar
where $j_l(x)$ denotes the spherical Bessel functions and
the cosine angle between the line-of-sight direction~$\Vang$ and the
pair separation vector~$\svec$ is $\mu=\Vang\cdot\hat\svec$.
With the distant-observer approximation, there are no ambiguities associated
with how to define the line-of-sight direction of the galaxy pair, as 
all angular directions are identical.

Due to the nonlocal nature of the power spectrum analysis, it is easier
to handle
the complex geometries of a given survey in measuring the correlation function.
However, its covariance matrix is highly correlated, and more importantly
measurements of the correlation function on large scales are difficult to
interpret due to the nontrivial integral constraints \citep{PEEBL80}.
By contraries, the power spectrum is free from the integral 
constraints, and its Fourier modes are independent.
Therefore, we will quantify the signal-to-noise ratios by using the 
redshift-space power spectrum, while we will also present our results in
configuration space.

A simplest unbiased estimator of the power spectrum in redshift space is
\beeq
\hat\PZ(\kvec)={1\over V_s}~\delta^s_{\kvec}~\delta^s_{-\kvec}-{1\over\bng}~,
\eneq
where $\delta^s_{\kvec}$ is the Fourier modes of the galaxy fluctuation in
redshift space and $V_s$ is the survey volume. To make the estimator unbiased,
we subtract the shot-noise contribution due to the discrete nature of galaxies.
The central limit theorem dictates that once many uncorrelated modes of 
the power spectrum estimates are added, the covariance matrix is well 
approximated by a Gaussian distribution. With the Gaussian approximation,
the covariance matrix can be straightforwardly computed as \citep{MEWH99}
\bear
\up{Cov}[\hat\PZ(\kvec)\hat\PZ(\kvec')]&&=
\left\langle\hat\PZ(\kvec)\hat\PZ(\kvec')\right\rangle
-\PZ(\kvec)\PZ(\kvec') \\
&&=(\delta_{\kvec,\kvec'}+\delta_{\kvec,-\kvec'})
\left[\PZ(k,\mu_k)+{1\over\bng}\right]^2 ~.\nonumber
\enar
Using equation~(\ref{eq:multi}), the covariance matrix of the multipole
power spectra is \citep{TANISA10}
\bear
\up{Cov}[\hat\MP_l(k)\hat\MP_{l'}(k')]&=&{(2l+1)(2l'+1)\over2}~\delta_{kk'}
\nonumber\\
&&\hspace{-70pt}\times\int_{-1}^1d\mu_k~\LL_l(\mu_k)\LL_{l'}(\mu_k)~
\left[\PZ(k,\mu_k)+{1\over\bng}\right]^2~.
\label{eq:covariance}
\enar
The covariance matrix is diagonal in Fourier modes~$k$, but is correlated
in angular multipoles~$l$. With the redshift-space power spectrum in 
equation~(\ref{eq:kaiser1}), the covariance matrix of the multipole power
spectra can be explicitly computed and is given in Appendix~\ref{app:cov}.

In the limit of highly biased objects ($\beta\RA0$), the redshift-space
power spectrum becomes independent of angle, and the covariance matrix
of the multipole power spectra become diagonal in angular multipole:
\beeq
\up{Cov}[\hat\MP_l(k)\hat\MP_{l'}(k')]=\delta_{kk'}\delta_{ll'}(2l+1)
\left[b^2\PM(k)+{1\over\bng}\right]^2~.
\label{eq:cov}
\eneq
Throughout the paper, we used equation~(\ref{eq:cov}) to compute the intrinsic
variance of the corresponding multipole power spectra and assumed that the
shot-noise contribution is negligible ($\bng\RA\infty$). Moreover,
the intrinsic variance of the multipole power spectra
can be further reduced by adding more measurements
at each Fourier mode. Given the survey volume~$V_s$ and the galaxy number 
density~$\bng$, the effective number of Fourier modes is 
(FKP: \citealt{FEKAPE94})
\beeq
N_k={1\over2}{4\pi k^2dk\over(2\pi)^3}\int dV_s\left(
{\bng\PZ\over1+\bng\PZ}\right)^2\simeq{1\over2}{4\pi k^2dk\over(2\pi)^3}~V_s~,
\label{eq:modes}
\eneq
where for simplicity we assumed that the galaxy sample is limited by the 
sample variance, not by shot-noise. The factor two in 
equation~(\ref{eq:modes}) arises due to the fact that the Fourier modes
represent a real quantity, i.e., the galaxy number density $n_g$.

\begin{figure}
\centerline{\psfig{file=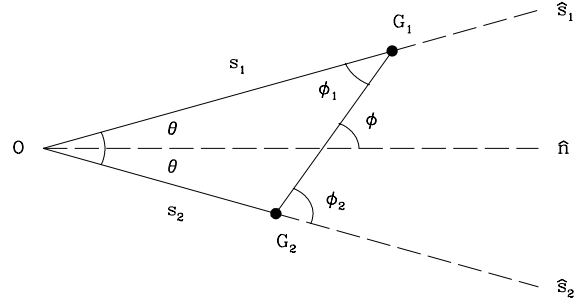, width=3.4in}}
\caption{Triangular configuration of the observer~$O$ and the galaxy 
pairs $G_1$ and $G_2$. The opening angle of the galaxy pair
is $\oo\equiv2\theta=\phi_2-\phi_1$, and the line-of-sight
direction of the pair is defined as the direction~$\Vang$ that bisects
the pair in angle, forming an angle $\phi=(\phi_1+\phi_2)/2$ with the
pair separation. With the distant-observer approximation,
three angles become identical $\phi=\phi_1=\phi_2$ ($\theta\RA0$).}
\label{fig:geometry}
\end{figure}

\subsection{Redshift-space distortion in wide angle surveys}
\label{ssec:wide}
The redshift-space correlation function without the distant-observer
approximation was first computed in \cite*{SZMALA98} by choosing a specific
coordinate system, in which their expansion in bipolar spherical harmonics
becomes particularly simple and there exist only a finite
number of terms. This calculation was extended to
non-flat universes \citep{MATSU00b}. \cite{SZAPU04} improved the calculation
of the redshift-space correlation function by expressing it in terms of
tri-polar spherical harmonics, in which the expansion coefficients depend
only on pair separation and only a finite number of those coefficients are
present. It was shown \citep{SZAPU04} that any coordinate system may be
chosen to compute the redshift-space correlation, but two including the choice
by \cite{SZMALA98} result in the most compact expression of the redshift-space
correlation function. Further extension of the calculation was made
first by \cite{PASZ08} to implement the selection function in the formula
and second
by \cite{MODU12} to account for the case where two galaxies of the pair
are separated at two different redshifts.

Drawing on these previous developments, we briefly review the formalism
for computing the redshift-space correlation function without the 
distant-observer approximation and discuss 
the deviation from the simple Kaiser formula with the distant-observer
approximation.
The full Kaiser formula for the redshift-space galaxy fluctuation in
equation~(\ref{eq:fullkaiser0}) can be recast in Fourier space as
\beeq
\ds=\int{d^3\kvec\over(2\pi)^3}~e^{i\kvec\cdot\svec}~
\left(b+f\mu_k^2
-i\mu_k{\alpha f\over k ~r}\right)\delta_m(\kvec)~,
\label{eq:fullkaiser0ft}
\eneq
and without adopting the distant-observer approximation the 
redshift-space correlation function can be formally written as
\bear
\label{eq:fullcorr}
\XZ&=&\int{d^3\kvec\over(2\pi)^3}~
e^{i\kvec\cdot(\svec_1-\svec_2)}\PM(k|z_1,z_2) \\
&\times&\left(b_1+f_1\mu_{\kvec_1}^2
-i\mu_{\kvec_1}{\alpha_1 f_1\over k ~r_1}\right)
\left(b_2+f_2\mu_{\kvec_2}^2
+i\mu_{\kvec_2}{\alpha_2 f_2\over k ~r_2}\right)~,\nonumber
\enar
where the quantities with subscript index represent those for each galaxy
of the pair, located at redshift $z_1$ and $z_2$ with angle $\hat\svec_1$ and
$\hat\svec_2$. It is noted that with two different line-of-sight directions
for each galaxy there are two different cosine angles 
$\mu_{\kvec_1}=\hat\svec_1\cdot\Kang$
and $\mu_{\kvec_2}=\hat\svec_2\cdot\Kang$, given a wavevector~$\kvec$. 

Moreover,
with two galaxies on the sky, the line-of-sight direction~$\Vang$ of 
the pair can be defined in various ways. Following the lead by
\cite{SZMALA98} and \cite{SZAPU04}, we define the line-of-sight direction
of the pair as the direction $\Vang$, bisecting the pair in angles 
(hence its cosine angle $\mu_{\kvec}=\Vang\cdot\Kang$ and
$\mu=\Vang\cdot\hat\svec$, 
see Fig.~\ref{fig:geometry} for the configuration).

In literature, the ``wide angle effects'' are often used to refer to the
deviation from the redshift-space power spectrum (the simple Kaiser formula)
in equation~(\ref{eq:kaiser1}) or the redshift-space correlation function
in equation~(\ref{eq:zcorr}). Comparing equation~(\ref{eq:fullcorr})
with equation~(\ref{eq:kaiser1}), the deviation can be attributed to 
two physically distinct parts: one involves the difference among
three cosine angles $\mu_{\kvec_1}$, $\mu_{\kvec_2}$, and 
$\mu_{\kvec}$, and the other arises from the additional velocity
contribution in the full Kaiser formula
that is proportional to the selection function~$\alpha$
in equation~(\ref{eq:fullkaiser0}).
The former can be legitimately referred to as the wide angle effect, since 
it represents the deviation from the one and only
line-of-sight direction in the
distant-observer approximation. Though the latter is often referred
to as the mode coupling (e.g., \cite{RASAPE10,RABEET12}), it just 
represents the additional velocity contribution, coupling the density and 
the velocity components, but leaving
each Fourier mode uncoupled. Certainly, the latter effect is independent
of how widely in angle galaxy pairs are separated. Here we refer to the 
above effects as the deviation from the simple Kaiser formula with the
distant-observer approximation.

Given the observed positions of the galaxy pair, $\svec_1=(\hat\svec_1$, $z_1$)
and $\svec_2=(\hat\svec_2$, $z_2$), 
the triangular configuration formed by the galaxy
pair and the observer is depicted in Fig.~\ref{fig:geometry}, and 
the redshift-space distances of the galaxy pairs and their pair
separation are related as
\bear
s&=&\left[s_1^2+s_2^2-2s_1s_2\cos\oo\right]^{1/2}~,\\
s_1&=&{\sin\phi_2\over\sin\oo}~s~, \\
s_2&=&{\sin\phi_1\over\sin\oo}~s~.
\enar
In \cite{SZAPU04}, the full redshift-space correlation
function in equation~(\ref{eq:fullcorr}) is expanded in terms of tri-polar
spherical harmonics, and it is evaluated in three different coordinate
systems. Following \cite{SZAPU04} and \cite{PASZ08}, we choose a coordinate
system, in which the triangle is confined in the $x$-$y$ plane and the pair
separation vector~$\svec$ is parallel to the $x$-axis.\footnote{The choice
of coordinate system is, however, a matter of preference, and the resulting
correlation function is independent of our choice.} With this choice of
coordinate system, the full redshift-space correlation function 
can be described by a finite number of terms that depend on
two angles $\phi_1$, $\phi_2$ and pair separation~$s$ as
\bear
&&\hspace{-20pt}
\XZ(s,\phi_1,\phi_2)=\sum_{i,j=0,1,2}
a_{ij}(s,\phi_1,\phi_2)\cos(i\phi_1)\cos(j\phi_2) \nonumber \\
&&+b_{ij}(s,\phi_1,\phi_2)\sin(i\phi_1)\sin(j\phi_2)~,
\label{eq:full}
\enar
where the coefficients non-vanishing under the distant-observer approximation
are
\bear
a_{00}&=&\left(b_1b_2+{b_2f_1+b_1f_2\over3}+{2f_1f_2\over15}\right)
\xi_0^2(s) \\
&& - \left({b_2f_1+b_1f_2\over6}+{2f_1f_2\over21}\right)\xi_2^2(s)
+{3f_1f_2\over140} ~\xi_4^2(s)~,  \nonumber \\
a_{20}&=& -\left({b_2f_1\over2}+{3f_1f_2\over14}\right) \xi_2^2(s) 
+{f_1f_2\over28} ~\xi_4^2(s) ~,  \\
a_{02}&=& -\left({b_1f_2\over2}+{3f_1f_2\over14}\right) \xi_2^2(s) + 
{f_1f_2\over28} ~\xi_4^2(s) ~,  \\
a_{22}&=&{f_1f_2\over15} ~\xi_0^2(s) - {f_1f_2 \over21}~\xi_2^2(s) 
+ {19f_1f_2\over140} ~\xi_4^2(s) ~, \\
b_{22}&=& {f_1f_2\over15}~\xi_0^2(s) -{f_1f_2 \over21}~\xi_2^2(s) - 
{4f_1f_2 \over35}~\xi_4^2(s) ~, 
\enar
the remaining coefficients are
\bear
\label{eq:coef}
a_{10}&=& \left(b_2 f_1+{2f_1f_2\over5}\right){\alpha_1\over r_1}~ \xi_1^1(s) 
 - {f_1f_2\over10}{\alpha_1\over r_1} ~\xi_3^1(s) ~,  \\
a_{01}&=& -\left(b_1 f_2+{2f_1f_2\over5}\right){\alpha_2\over r_2} ~\xi_1^1(s) 
 + {f_1f_2\over10}{\alpha_2\over r_2}~ \xi_3^1(s) ~,  \\
a_{11}&=& {f_1f_2\over3}{\alpha_1 \alpha_2 \over r_1r_2}~\xi_0^0(s) 
 - {2f_1f_2\over 3}{\alpha_1 \alpha_2\over r_1r_2}~ \xi_2^0(s)~, \\
a_{21}&=& -{f_1f_2\over5}{\alpha_2\over r_2}~\xi_1^1(s) + 
{3f_1f_2\over10}{\alpha_2\over r_2}~ \xi_3^1(s) ~,  \\
a_{12}&=& {f_1f_2\over5}{\alpha_1\over r_1}~\xi_1^1(s) -
{3f_1f_2\over10}{\alpha_1\over r_1}~ \xi_3^1(s) ~, \\
b_{11}&=&{f_1f_2\over3}{\alpha_1 \alpha_2 \over r_1r_2} ~\xi_0^0(s)
+ {f_1f_2\over3}{\alpha_1 \alpha_2\over r_1r_2} ~\xi_2^0(s) ~,  \\
b_{21}&=& -{f_1f_2\over5}{\alpha_2\over r_2} ~\xi_1^1(s) - 
{f_1f_2\over5}{\alpha_2\over r_2}~ \xi_3^1(s) ~,  \\
\label{eq:b12}
b_{12}&=& {f_1f_2\over5}{\alpha_1\over r_1} ~\xi_1^1(s)+
{f_1f_2\over5}{\alpha_1\over r_1}~ \xi_3^1(s) ~,
\enar
and we defined
\beeq
\xi_l^n(x)=\int{dk\over2\pi^2}~k^nj_l(kx)\PM(k|z_1,z_2)~,
\eneq
where $\PM(k|z_1,z_2)=D(z_1)D(z_2)\PM(k|z=0)$.
These coefficients are derived by extending the \cite{PASZ08} calculations
to the case with two different bias factors and selection functions
\citep{MODU12}. In this expansion,  
the redshift-space correlation function with the distant-observer approximation
in equation~(\ref{eq:zcorr}) is 
\bear
\label{eq:simple}
\XZ(s,\phi)&=&a_{00}+a_{02}\cos2\phi+a_{20}\cos2\phi \nonumber\\
&&+a_{22}\cos^2(2\phi)+b_{22}\sin^2(2\phi)~.
\enar

\begin{figure}
\centerline{\psfig{file=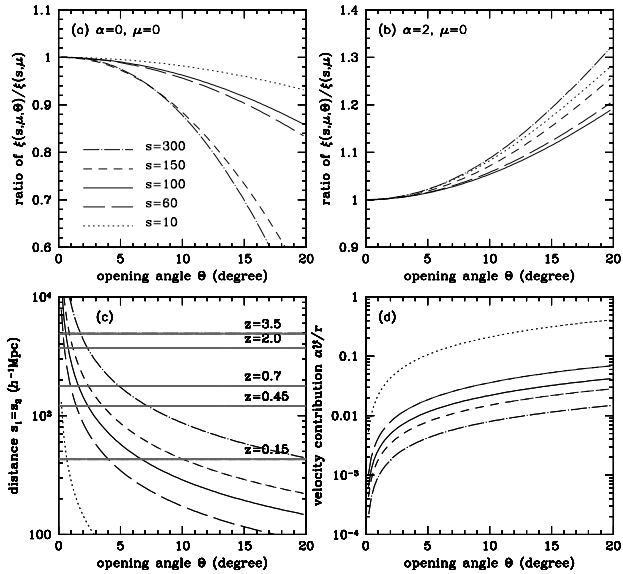, width=3.4in}}
\caption{Full redshift-space correlation function in galaxy surveys. 
The full redshift-space correlation function in equation~(\ref{eq:fullcorr}) 
depends on the triangular configuration $(s,\mu,\oo)$ formed by the galaxy 
pair and the observer in Fig.~\ref{fig:geometry}. For illustration, 
we consider equilateral
triangular shapes (i.e., $s_1=s_2$ and $\mu=0$). Upper panels show 
the ratio of the full correlation function to the correlation
function computed by using the distant-observer approximation in 
equation~(\ref{eq:zcorr}) at the redshift of the galaxy pair ($z_1=z_2$).
The deviation therefrom arises due to the wide angle effect and 
the velocity contribution.
($a$)~The selection function is set ~$\alpha=0$, and there is no velocity
contribution, representing the deviation purely due to the wide angle effect.
($b$)~The velocity contribution is considered ($\alpha=2$).
($c$)~The distance from the observer to each galaxy of the pair.
Horizontal lines show the distance to the redshifts in the label. For 
reference, the SDSS LRG covers the redshift range $z=0.15\sim0.45$
(Euclid: $z=0.7\sim2.0$ and the BigBOSS: $z=2.2\sim3.5$).
Most galaxy pairs measured in these surveys will have small opening angles. 
($d$)~Approximate velocity contribution $\alpha\VV/r$ with 
$V=10^{-3}$ ($300~\kms$)
to the full Kaiser formula for the redshift-space fluctuation~$\ds$ in 
equation~(\ref{eq:fullkaiser0}). For illustration, we assumed
a uniform galaxy sample with $b=1$ (or dark matter).
Uniform galaxy samples with higher bias factor would further 
reduce the deviation from the distant-observer approximation.}
\label{fig:wide}
\end{figure}

Figure~\ref{fig:wide} compares the full redshift-space correlation function
in equation~(\ref{eq:fullcorr}) to the redshift-space correlation function with
the distant-observer approximation in equation~(\ref{eq:zcorr}).
They are computed by using equations~(\ref{eq:full}) and~(\ref{eq:simple}),
respectively.
While the latter is independent of the opening angle~$\oo$,
the former depends
on the triangular configuration characterised by~$\oo$ as well as
the pair separation~$s$ and the cosine angle $\mu=\cos\phi$. The upper panels
show the deviation from the simple Kaiser formula with the distant-observer
approximation. Panel~($a$) illustrates the deviation due to the wide angle
effect, in which the selection function is arbitrarily set zero $\alpha=0$
and hence there is no velocity contribution.
The deviation from the distant-observer approximation naturally becomes
substantial as the opening angle increases, but it remains small at
$\oo\leq5$~degrees.
With the feature in the
correlation function around the BAO scale, the ratio is {\it not}
a simple scaling of pair separation~$s$. Panel~($b$) shows the full deviation
from the simple Kaiser formula, including the velocity contribution.
Compared to Fig.~\ref{fig:wide}$a$, it is apparent that the velocity
contribution in this case is non-negligible, as we explain below.
In \cite{RASAPE10}, they computed the wide angle effect
in various triangular configurations and tested the accuracy of the wide angle
formula against numerical simulations.

While the velocity contribution is independent of how widely galaxy pairs 
are separated in angle, the angular separation affects the velocity 
contribution by changing the legs of the triangle. Panel~($c$)
shows the distance from the observer to each galaxy
of the pair as a function of opening angle. With a fixed pair separation
and a cosine angle, the galaxy pairs are closer to the observer as the opening
angle increases. For example, the pair galaxies (solid) with $s=100\hmpc$ 
in the equilateral configuration are about $300\hmpc$ away from the observer 
to have an opening angle of $\oo=10$~degrees. The distances to the closer 
galaxy of the pairs are even closer for any triangular configuration with 
$\mu\neq0$. The horizontal lines  indicate redshifts at various
distances, illustrating typical distances to galaxies measured in galaxy
surveys. To achieve a certain number of galaxies measured in each survey,
galaxy surveys are designed to cover a large cosmological volume, and 
galaxies in those surveys are inevitably far away from the observer.
Beyond the SDSS, galaxies in surveys like Euclid and the BigBOSS will be
at least $1~\hgpc$ away from the observer. Consequently, the opening angles 
of galaxy pairs in those surveys will be a few degrees at maximum
(pairs with largest separation~$s$), and in most cases
the opening angles of galaxy pairs will be close to zero.

Panel~($d$) shows the approximate velocity contribution term $\alpha\VV/r$
of the full Kaiser formula $\ds$ for the redshift-space galaxy fluctuation
in equation~(\ref{eq:fullkaiser0}). For illustration, we assumed 
a typical line-of-sight velocity of $V=10^{-3}$ ($300~\kms$).
As the opening angle
increases, the distance to galaxies decreases substantially, and this in turn
increases the velocity contribution. The inverse scaling with the distance
to galaxies arises due to the distortion in volume: The volume element is
proportional to $r^2$ in the mean, and its perturbation is therefore
$2~\delta r/r\sim2~V/r$ with the peculiar velocity being 
the leading contribution to the radial distortion 
$\delta r\simeq|s-r|$. 
At a typical opening angle
$\oo\ll1$~degree, the velocity contribution is negligible, and its contribution
to the correlation function is even smaller.

\subsection{Connection to the relativistic formula}
\label{ssec:gr}
The general relativistic description of galaxy clustering has been developed
in the past few years \citep{YOFIZA09,YOO10} (see also 
\citealt{BODU11,CHLE11,JESCHI12}).
The relativistic description of galaxy clustering follows the same principle
as in the redshift-space distortion: The observed number of galaxies is 
conserved when expressed in terms of the physical and the observed quantities 
\citep{YOO09}. In the context of redshift-space distortion, the physical 
quantities represent the real-space quantities, and the observed quantities
represent the redshift-space quantities. However, the distinction in the 
physical and observed quantities is more general than in the redshift-space
distortion case, and its understanding reveals the subtlety of those quantities
associated with gauge issues \citep{YOFIZA09,YOO10}.

Observed quantities such as the observed redshift are related to physical
quantities of galaxies at their rest frame by the photon geodesic equation,
and perturbations along the photon path result in distortion in the observed
quantities. The dominant contribution to the distortion in the observed
redshift is the peculiar velocity, but there exist other relativistic
contributions such as the Sachs-Wolfe and the integrated Sachs-Wolfe
effects \citep{SAWO67}. Similarly, other velocity and relativistic 
contributions are present in the observed galaxy fluctuation due to the
distortion in volume between the physical and the observed 
\citep{YOFIZA09,YOO10}.

In addition to the velocity contribution from the volume distortion,
another velocity term arises in conjunction with the evolution of the source
galaxy number density~$\bng$, characterised by the evolution factor
\beeq
e=3+{d\ln\bng\over d\ln(1+z)}~,
\label{eq:evol}
\eneq
where the factor three arises due to the volume dilution and
may be absorbed into~$\bng$ by redefining it as the physical
number density. In redshift space, the redshift-space distance~$s$
described by the observed redshift is different from the real-space 
distance~$r$, and with the evolving galaxy population this mismatch 
gives rise to additional velocity contribution. In the conservation 
equation~(\ref{eq:conr}), this contribution is represented by the ratio
of the mean number densities $\bng(r)/\bng(s)$ in real-space and 
redshift-space,
and it is related to the selection function in the full Kaiser formula in 
equation~(\ref{eq:fullkaiser0}) as
\beeq
\alpha={d\ln r^2\bng\over d\ln r}=2+{rH\over 1+z}(e-3)~.
\eneq
Apparent in its definition, the selection function~$\alpha$
contains two physically distinct velocity contributions, by which
the factor two represents the volume distortion and is independent
of the evolving galaxy population, while the evolution factor~$e$
represents solely the distortion due to the source population.
It is now evident that setting $\alpha=0$ requires a very strange evolution
of the galaxy population.
A complete description of the connection between the Kaiser formula and
the relativistic formula and its related gauge issues are given in 
\cite{YOHAET12}.

In \cite{YOHAET12}, we showed that while the Kaiser formula is devoid of any 
relativistic contribution, it properly reproduces the velocity contribution
of the relativistic formula, except a missing velocity contribution 
from the fluctuation in the luminosity distance. Without it, the Kaiser formula
can only describe galaxy samples without any selection bias, except one
from the observed redshift.  If galaxy samples are selected by its rest-frame
luminosity, it has additional velocity contribution that is proportional to
the luminosity function slope~$p$ at the threshold, and it can be implemented
to the full Kaiser formula by replacing the selection function
\beeq
\alpha\RA\af\equiv\alpha+5p~(\HH r-1)~,
\label{eq:correc0}
\eneq
where the luminosity function slope is
\beeq
p={d\log\bng\over dM}=-0.4~{d\log\bng\over d\log L}~,
\eneq
and the absolute magnitude and luminosity are~$M$ and~$L$, respectively

Furthermore, the mapping between the real-space and the redshift-space 
takes place through the past
light cone, as we observe galaxies by measuring photons.
In other words, galaxies at higher redshift
are not only farther away from the observer, but also farther back in time.
This implies the derivative term in the full Kaiser formula for the
redshift-space
galaxy fluctuation in equation~(\ref{eq:fullkaiser0}) is the total derivative
along the photon path:
\beeq
{d\over dr}={\partial\over\partial r}-{\partial\over\partial\tau}~,
\eneq
where $\tau$ is the conformal time. It is only with this relation that the
full Kaiser formula in equation~(\ref{eq:fullkaiser0}) reproduces the complete
velocity terms in the relativistic formula. With the additional
correction in equation~(\ref{eq:correc0}) due to the fluctuation in the
luminosity distance, the full Kaiser formula for the redshift-space galaxy 
fluctuation is \citep{YOHAET12}
\bear
\label{eq:fullkaiser1}
\delta_z&=&b~\dm-\left({d\over dr}+{\af\over r}\right)\VV \nonumber\\
&=&b~\dm-{1+z\over H}{dV\over dr}-e~V+2~V-{2V\over\HH r}\nonumber\\
&&+{1+z\over H}{dH\over dz}~V-5p~\left(1-{1\over\HH r}\right)~V~.
\enar
Again, we emphasise that in addition to the velocity contribution in 
equation~(\ref{eq:fullkaiser1}) there exist additional relativistic 
contributions \citep{YOFIZA09,YOO10,YOHAET12}. The relativistic contributions
are smaller than the velocity contributions, and they are omitted here
for simplicity.

The observation that the derivative from the Jacobian in
equation~(\ref{eq:conr}) is the total derivative reveals the 
mistakes made in the Kaiser formulae in equations~(\ref{eq:kaiser0}) 
and~(\ref{eq:fullkaiser0ft}) in Fourier space and hence its correlation 
function in equation~(\ref{eq:fullcorr}). In all those cases, the derivative
term is regarded as a partial derivative in space, which is sensible
in equation~(\ref{eq:kaiser0}) when only the leading contribution 
$\sim\mathcal{O}(\delta)$ is considered. However, when the velocity terms
are considered as in equations~(\ref{eq:fullkaiser0ft}) 
and~(\ref{eq:fullcorr}), the time derivative should be considered in order
to fully recover the velocity terms in equation~(\ref{eq:fullkaiser1}):
\bear
-{d\VV\over dr}&=&-{1+z\over H}{dV\over dr}-V+{1+z\over H}{dH\over dz}~V\\
&=&\int{d^3\kvec\over(2\pi)^3}
\left[f\mu_k^2\delta_{\kvec}-{i\mu_k\over\HH}\left(v'_\kvec-\HH v_\kvec
+{dH\over dz}~v_\kvec\right)\right]e^{i\kvec\cdot\svec}~,\nonumber
\enar
where $V_\kvec=-i\mu_k v_\kvec$. 
Consequently, the full redshift-space correlation function in 
equation~(\ref{eq:fullcorr}) and the following 
equations~(\ref{eq:coef})$-$(\ref{eq:b12}) are affected by the missing
velocity contributions from the time derivative. Therefore, it is noted
that the velocity contribution is in fact non-vanishing, even when $\af=0$ 
(or $\alpha=0$) or the distant-observer approximation is adopted. 
These corrections can be made to those equations
in Fourier space, simply by replacing the selection function 
\beeq
\af\RA\af+\HH r-r~{dH\over dz}~,
\label{eq:correc2}
\eneq 
where the time derivative $v'_{\kvec}$ 
of velocity is ignored, consistently as other
gravitational potential contributions. Note, however, that
while equation~(\ref{eq:correc2}) accounts for the fact that 
equations~(\ref{eq:kaiser0}) and~(\ref{eq:fullkaiser0ft})
(and hence the coefficients in Eqs.~[\ref{eq:coef}]$-$[\ref{eq:b12}])
incorrectly compute the total derivative in Fourier space, 
the selection function~$\af$ in equation~(\ref{eq:correc0}) is correct.

\section{result}
\label{sec:result}

\subsection{Survey geometry}
\label{ssec:geometry}
Having established the formulae for computing the redshift-space galaxy
clustering and its associated errorbars, we are now in a position to quantify 
the deviation of galaxy clustering measurements in galaxy surveys
from the simple Kaiser formula with the distant-observer approximation.
Here we consider the simplest survey geometry, in which the survey area
is a single contiguous region, fully characterised by its sky coverage $\fsky$ 
and redshift range. For simplicity, we assume that 
the angular selection function is unity
within the survey region, and there are no holes in the survey region.
In practice, survey regions are a sum of disconnected patches on the sky 
(e.g., \citealt{YOADET00,CODAET01}). 
Moreover, each patch often contains numerous holes and is
described by nontrivial angular selection function, all of which 
discourage observers from measuring galaxy clustering at a large separation
(see, however, \citealt{LYNDE71} for likelihood methods in these cases).
Surveys are often designed to have 
contiguous patches of size at best somewhat larger than the BAO scale (e.g., 
\citealt{YOADET00,SCBLET07,BLBRET11}), but significantly larger patches
are in practice difficult to construct in the current generation of surveys.
Therefore, our simplified geometry will maximise the deviation
from the simple Kaiser formula with the 
distant-observer approximation by allowing widely separated galaxy pairs 
to be used for measuring galaxy clustering that are often unavailable
in realistic galaxy surveys.

\begin{figure}
\centerline{\psfig{file=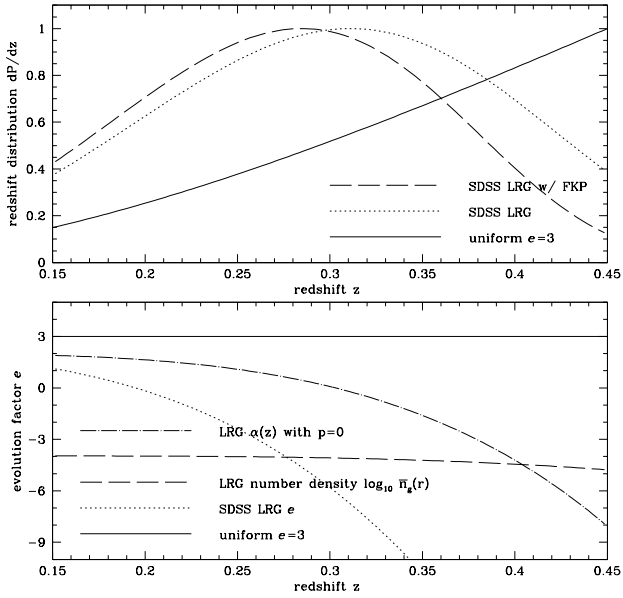, width=3.4in}}
\caption{Redshift distribution of the SDSS galaxy sample and the evolution
factor~$e$ of its number density. In the upper panel, various curves show the 
normalised redshift distributions (or the radial selection function) 
in equation~(\ref{eq:pzz}) 
for a galaxy sample with a constant comoving number density
($e=3$: solid) and the SDSS LRG sample (dotted). With the FKP weighting
in equation~(\ref{eq:FKP}), 
the redshift distribution (dashed) is shifted to lower
redshift and is closer to the uniform (solid).
The bottom panel shows the
evolution factor~$e$ of the galaxy number density. By definition, the uniform
sample has $e=3$, diluting only due to the volume expansion. The dotted curve
represents the SDSS LRG sample, indicating that its number density evolves 
rapidly in time 
(the comoving number density of the SDSS LRG sample is shown as dashed,
and the corresponding selection function~$\alpha$ is shown as dot-dashed).
The galaxy samples in future surveys are assumed to have a uniform distribution
($e=3$: solid), but with different redshift ranges.}
\label{fig:zdist}
\end{figure}

The radial selection function is the expected number density of galaxies,
or the unclustered mean number density. In general, the mean galaxy number
density $\bng(z)$ is obtained by averaging the observed galaxy number density
over the survey area
given redshift bins, and hence it is subject to the
sample variance error. Further complication arises due to the way the galaxy 
sample is defined. For example, if galaxies are selected based on luminosity,
proper treatments of $K$-correction and $E$-correction are required to 
ensure that galaxies are selected by the same luminosity threshold at their
rest frame and galaxies in the sample are an identical population over the
redshift range. So as to obtain physical insight, we consider idealised 
situations, in which the comoving galaxy number density is constant ($e=3$
in Eq.~[\ref{eq:evol}]) and the radial selection function is constant 
($\alpha=2$). 
We refer to these samples as ``uniform'' galaxy samples.\footnote{In 
literature, a volume-limited galaxy sample is often synonymously
used as a uniform sample. However, a volume-limited sample refers to
a galaxy sample constructed by imposing a constant luminosity threshold
at each redshift (and hence different threshold in observed
flux at each redshift), while a uniform galaxy sample simply refers to
a galaxy sample with a constant comoving number density. A volume-limited
sample can be a uniform sample, but in principle they are unrelated.}

\begin{figure}
\centerline{\psfig{file=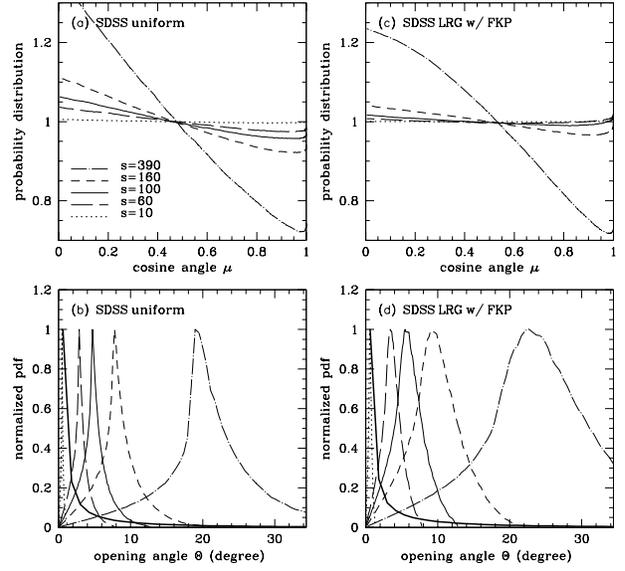, width=3.4in}}
\caption{Probability distribution of triangle shapes formed by SDSS galaxy
pairs and the observer. Pair distributions are affected by the number density
evolution and the survey geometry. For simplicity, the survey geometry is
defined in terms of redshift range and sky coverage only; no holes or disjoint
regions in the survey areas are assumed. Upper panels show the distribution of 
the cosine angle between the pair separation vector $\svec$
and the line-of-sight direction $\Vang$ bisecting the pair in angle.
Bottom panels show the distribution of the opening angle of galaxy pairs
seen by the observer at origin. Various curves represent different
separation length $s=|\svec|$ in units of $\hmpc$. Thick solid curves
show the distribution of the opening angle, averaged over all galaxies
with any pair separations. With the FKP weighting, more weight is given
to galaxies at lower redshift.}
\label{fig:adist}
\end{figure}

We consider three galaxy surveys with the simplified geometry: the 
SDSS, Euclid,
and the BigBOSS. While the wide angle effect in the SDSS is already discussed
at length in \cite{SAPERA12} with correct survey geometry, we include the SDSS
to set the stage for our calculations of the wide angle effect in future 
surveys, as the 
wide angle effect is largest in the SDSS among the surveys we consider here.

Figure~\ref{fig:zdist} describes the redshift distribution of the SDSS galaxy
sample and the evolution factor~$e$ in equation~(\ref{eq:evol}). 
The redshift distribution
$\PP(z)$ is simply the number of expected galaxies at a given redshift bin
and is related to the comoving galaxy number density as
\beeq
\PP(z)={4\pi\fsky\over N_\up{tot}}~{r^2\over H}~\bng~,
\label{eq:pzz}
\eneq
where $N_\up{tot}$ is the total number of observed galaxies, ensuring that
the redshift distribution is properly normalised. 
The redshift distribution of 
a uniform galaxy sample (solid) in Fig.~\ref{fig:zdist} is skewed to higher
redshift due to a larger volume at high redshift. The dotted curve shows
the approximate redshift distribution of the SDSS LRG sample 
\citep{EIANET01,COEIET08},
and the dashed curve shows the redshift distribution with further FKP
weighting \citep{FEKAPE94}
\beeq
w(r)\propto{\bng\over1+\bng P}~,
\label{eq:FKP}
\eneq
where $P$ is the power spectrum at a scale of interest, but is often assumed
to be a constant. We adopt $P=10^4~(\hmpc)^3$ (e.g., \citealt{PEREET10}).

The bottom panel shows the evolution factor~$e$ in equation~(\ref{eq:evol}) 
and the SDSS LRG number density $\bng(z)$. 
The redshift distribution is related to the evolution
factor (and hence the selection function) as
\bear
{d\ln\PP\over d\ln(1+z)}&=&2~{1+z\over rH}-{1+z\over H}{dH\over dz}
+(e-3) \nonumber \\
&=&{1+z\over rH}\left(\alpha-r {dH\over dz}\right)~.
\enar
By construction, a uniform galaxy sample
is of $e=3$ (solid). The dotted curve shows the evolution factor~$e$ of
the SDSS LRG sample (or the dot-dashed curve for the selection 
function~$\alpha$), 
illustrating that non-trivial velocity contributions 
are present in galaxy clustering due to
the evolving population. The mean number density (dashed) of the SDSS LRG 
sample is about a factor ten lower at $z=0.45$ than at the lower boundary
in redshift. For Euclid and the BigBOSS, we assume uniform galaxy samples
(solid curves) but with different redshift ranges.

\begin{figure}
\centerline{\psfig{file=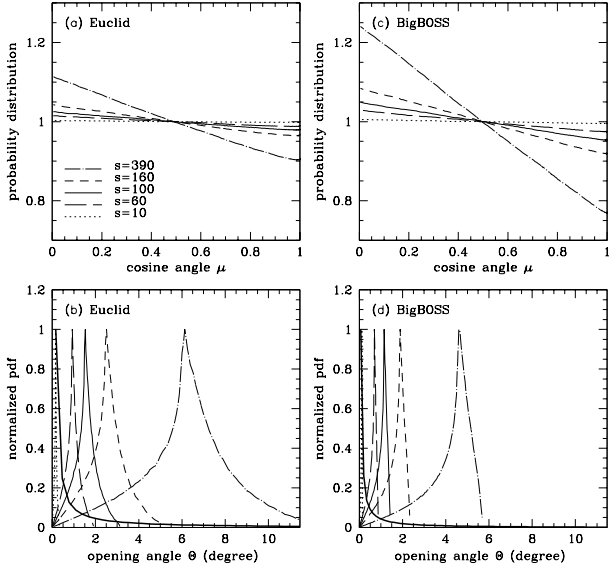, width=3.4in}}
\caption{Probability distribution of triangle shapes formed by galaxy pairs
and the observer in Euclid and the BigBOSS,
in the same format as Fig.~\ref{fig:adist}. It is assumed that the sky coverage
of Euclid is a half ($\fsky=1/2$) and the redshift range is $z=0.7\sim2.0$.
The BigBOSS is assumed to cover 14,000 deg$^2$ ($\fsky=0.34$) at
$z=2.2\sim3.5$. Given the mean distance to galaxy pairs, typical
opening angles in these surveys are a lot smaller than in the SDSS.
However, the non-uniform distribution of~$\mu$ is still present, as it is
inherent to survey geometry.}
\label{fig:adist2}
\end{figure}

To quantify the probability distribution of triangle shapes in three galaxy
surveys, we create mock catalogues of galaxies
by populating the survey region with random particles, and tabulate
the probability distribution as a function of triangular configuration
($s$, $\mu$, $\oo$) by counting pairs. 
Figures~\ref{fig:adist} and~\ref{fig:adist2}
describes the probability distribution of the
cosine angle~$\mu$ and the opening angle~$\oo$ of the galaxy samples
in the SDSS, Euclid, and the BigBOSS.
The upper panels show the 
distribution of the cosine angle at a given pair separation,
averaged over all pairs with various opening angles.
In an idealised situation,
the cosine angle of pair separations should be uniformly distributed,
as it is nearly so for pairs with small separations (dotted). 
However, even for a uniform sample,
there exist somewhat more pairs along the transverse direction than along
the line-of-sight direction simply due to the difference in volume of pairs
\bear
N_\up{pair}^\parallel&\propto&\int dr\int d^2\Vang~{1\over2}
\left[\left(r+{s\over2}\right)^2+\left(r-{s\over2}\right)^2\right]~,\\
N_\up{pair}^\perp&\propto&\int dr\int d^2\Vang~r^2\sin(\theta_{\Vang}+\oo)~.
\enar
Therefore, the fractional deviation to the leading order in $\oo$ and $s/r$ is
\beeq
{\Delta N_\up{pair}^\parallel\over N_\up{pair}^\perp}\simeq-{3\over2}~s~
{\sin\theta_M\over1-\cos\theta_M}~{r^2_\up{max}-r^2_\up{min}
\over r^3_\up{max}-r^3_\up{min}}\sim-
{s\over r_\up{avg}}\sqrt{1-\fsky\over\fsky}~,
\eneq
where $\theta_M$ is the maximum pair separation in angle
and $(r_\up{max},r_\up{min})$ are the radial distances from the observer,
 specifying the survey boundary.

The cosine angle~$\mu$
between the pair separation and the line-of-sight direction is affected
by the sky coverage of each survey (including the presence of disjoint regions 
and holes) and is largely 
independent of distance to galaxies from the observer.
The deviation from the uniform distribution of the cosine
angle is proportional to the pair separation, but it also depends on
the sky coverage. At larger separation (dot-dashed), the boundary 
effect becomes more important in determining
the probability distribution, as certain
triangular shapes cannot fit in the geometry.
Consequently, a proper weight should be given to account for the non-uniform
distribution of the cosine angle, when measuring the multipole
correlation functions or the multipole power spectra \citep{KASABL12}.
Compared to the probability distribution in the SDSS in Fig.~\ref{fig:adist}, 
the $\mu$-distribution in Fig.~\ref{fig:adist2}
is more uniform at the same separation in Euclid and the BigBoss, 
reflecting the difference in their sky coverage (Euclid: $\fsky=0.5$,
BigBOSS: $\fsky=0.34$, SDSS: $\fsky=0.25$).

\begin{figure}
\centerline{\psfig{file=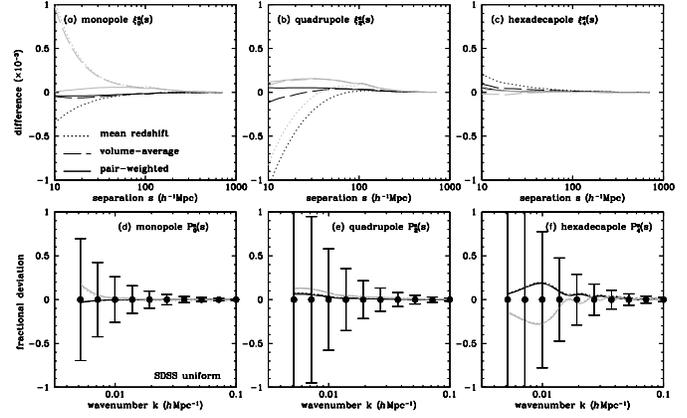, width=2.5in, angle=-90}}
\caption{Systematic errors of the simple Kaiser formula with the
distant-observer approximation for the SDSS uniform galaxy sample. 
Gray curves in each panel represent those for the SDSS LRG with the FKP
weighting.
Top rows show the difference in the multipole correlation functions $\MX_l(s)$,
and bottom rows show the fractional deviation in
the multipole power spectra $\MP_l(k)$. Given a pair separation~$s$ and a
cosine angle~$\mu$, the correlation function is obtained by averaging over 
all galaxy pairs with different opening angles~$\oo$ that fit in the SDSS 
geometry described in Section~\ref{ssec:geometry} (Eq.~[\ref{eq:fulld}]).
The multipole correlation
functions are obtained by accounting for the non-uniform distribution of the
cosine angle~$\mu$ in Fig.~\ref{fig:adist}, and the multipole power spectra
are obtained by Fourier transforming the corresponding multipole correlation
functions (see Eq.~[\ref{eq:mulp}]). Compared to the full redshift-space
correlation function, the fractional deviation of the simple Kaiser formula
with the distant-observer approximation is obtained in three different ways
of evaluation (see text for details): mean-redshift (dotted), volume-average
(dashed), and pair-weighted (solid).
{\it Upper panels:} Difference in the multipole correlation
functions $\MX_0(s)$, $\MX_2(s)$, and $\MX_4(s)$.
{\it Bottom panels:}~Fractional deviation of the multipole power spectra
$\MP_0(k)$, $\MP_2(k)$, and $\MP_4(k)$.
The difference in various curves is largely obscured, as the scales 
in the bottom panels are vastly different from those in the upper panels.
The measurement uncertainties are computed by using
the covariance matrix in equation~(\ref{eq:cov}), valid in the limit of
highly biased objects with negligible shot-noise contribution. The measurement
uncertainties are in practice larger and weakly correlated.}
\label{fig:powSDSS}
\end{figure}

The bottom panels show the distribution of the opening angle~$\oo$ at a given
pair separation, averaged over all galaxies with various cosine angle. 
The distribution of the opening angle~$\oo$ 
is mainly affected by distances to those
galaxies from the observer, but it also depends on the 
sky coverage of each survey. For a galaxy
sample with isotropic distribution, the mean cosine angle is $\bar\mu=1/2$,
and this configuration, placed at a distance~$r$, would yield an opening angle
$\bar\oo\simeq0.5~s/r$, when $\oo\ll1$. Naturally, the opening angle 
is close to zero for pairs with small separations (dotted), while it peaks
at larger angle for pairs with larger separations (dot-dashed). 

The change of the $\oo$-distribution due to the FKP weighting 
in Fig.~\ref{fig:adist} is to
broaden the distribution with a slight shift to a larger opening angle,
since its effect is largely to pull individual galaxies to lower redshift.
Thick solid curves in the bottom panels
show the distribution of the opening angle, averaged over all galaxies,
not restricted to pair separations.
With fewer galaxies at large separation, the average opening angle of galaxy
pairs in the SDSS is close to zero \citep{OKMAET08,BEBLET11,SAPERA12}.
Compared to the probability distribution in the SDSS
\citep{SAPERA12}, our probability distribution function in Fig.~\ref{fig:adist}
prefers slightly larger opening angles due to the idealised geometry.
The $\oo$-distribution in Fig.~\ref{fig:adist2}
is much smaller than in the SDSS, 
primarily because galaxy pairs in Euclid and the BigBOSS are farther away 
by a factor of few on average. For pair separations we considered in 
Fig.~\ref{fig:adist2}, the impact of the sky coverage on the $\oo$-distribution
is rather weak in Euclid and the BigBOSS. However, the sky coverage
or the presence of disjoint regions in practice puts significant constraints
on the probability distribution, further reducing the opening angles,
as galaxies in two disjoint regions are not used for computing the correlation
function.

\begin{figure}
\centerline{\psfig{file=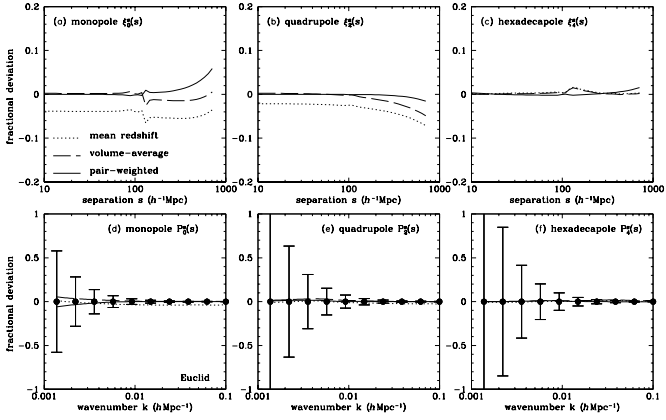, width=2.5in, angle=-90}}
\caption{Systematic errors of the simple Kaiser formula with the
distant-observer approximation in Euclid. Each panel shows the fractional
deviation in the multipole correlation function~$\MX_l(s)$ and the
multipole power spectra~$\MP_l(k)$. A small but sudden change in $\MX_0(s)$
arises  at $s\sim130\hmpc$ when the monopole changes its sign to negative,
crossing zero. The survey volume is $V_s=96$ 
$(\hgpc)^3$, approximately
50 times larger than in the SDSS. Correspondingly, the errorbars and the 
minimum wavenumber are smaller. Galaxies in Euclid are  farther
away than in the SDSS. The correlation functions with the distant-observer
approximation (dotted) at the mean redshift are off in amplitude,
while their shape is still consistent.}
\label{fig:powEUCL}
\end{figure}

\subsection{Deviation from the simple Kaiser formula with
the distant-observer approximation}
With the full probability distribution of the triangular configuration
formed by galaxy pairs and the observer,
we are now ready to compute the full redshift-space correlation
function, accounting for the deviation from the simple Kaiser formula with the
distant-observer approximation. At a given separation~$s$ and a cosine
angle~$\mu$, there exist a number of galaxy pairs with various opening
angle, shown in the bottom panels of Figs.~\ref{fig:adist} 
and~\ref{fig:adist2}.
Galaxy pairs that are closer to the observer have larger opening angles
at fixed values of~$s$ and~$\mu$. For each opening angle~$\oo$
available in a given survey region, we use equation~(\ref{eq:full}) to compute
the full redshift-space correlation function $\XZ(s,\phi_1,\phi_2)$,
where two angles $\phi_1$ and $\phi_2$ between the pair separation 
vector~$\svec$ and two different line-of-sight directions $\Xang_1$ and
$\Xang_2$ are related to the opening angle~$\oo$ and the cosine 
angle~$\mu$ as
\bear
\mu&=&\cos\left({\phi_1+\phi_2\over2}\right)~,\\
\oo&=&\phi_2-\phi_1~.
\enar
The triangular configuration is fully specified either by $(s,\mu,\oo)$ or
by $(s,\phi_1,\phi_2)$.
The resulting correlation function is then
averaged over all galaxy pairs with the
same pair separation~$s$ and the cosine angle~$\mu$ as
\beeq
\XZ(s,\mu)={\int d\oo~\pdf(\oo,\mu,s)~\XZ(s,\phi_1,\phi_2)
\over\int d\oo~\pdf(\oo,\mu,s)}~,
\label{eq:fulld}
\eneq
where the probability distribution $\pdf(\oo,\mu,s)$ in the bottom panel
of Figs.~\ref{fig:adist} and~\ref{fig:adist2}
is proportional to the number of galaxy pairs
at a given triangular configuration.
We then decompose the resulting correlation function into the 
redshift-space multipole correlation functions by using
\beeq
\MX_l(s)={2l+1\over2}\int_{-1}^1d\mu~\LL_l(\mu)~\XZ(s,\mu)~.
\label{eq:mxl}
\eneq
The non-uniform distribution of the cosine angle~$\mu$ in Figs.~\ref{fig:adist}
and~\ref{fig:adist2}
prevents from using a simple approach to obtaining the multipole correlation
functions:
\beeq
\MX_l(s)\neq{2l+1\over2}{\int d\mu~\LL_l(\mu)\int d\oo~\pdf(\oo,\mu,s)
~\XZ(s,\phi_1,\phi_2) \over\int d\mu\int d\oo~\pdf(\oo,\mu,s)}~.
\label{eq:mxl2}
\eneq

\begin{figure}
\centerline{\psfig{file=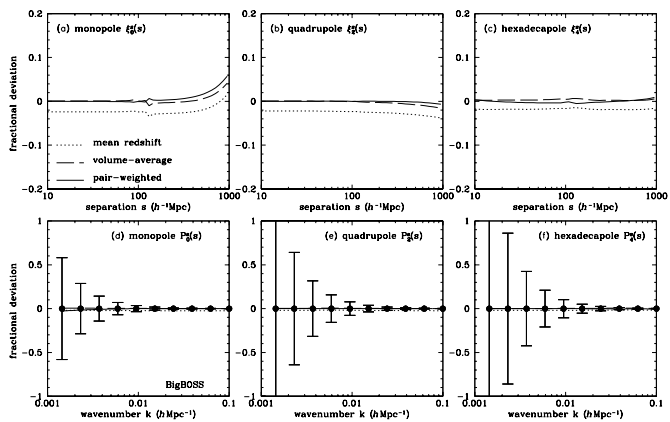, width=2.5in, angle=-90}}
\caption{Systematic errors of the simple Kaiser formula with the
distant-observer approximation in the BigBOSS, in the same format as 
Fig.~\ref{fig:powEUCL}. The survey is narrower but deeper than Euclid, yielding
$V_s=82$ $(\hgpc)^3$, similar to Euclid. The Kaiser formula
is accurate in all practical senses.}
\label{fig:powBIGB}
\end{figure}

Top panels in Figs.~\ref{fig:powSDSS}$-$\ref{fig:powBIGB}
show the difference of the multipole  correlation functions~$\MX_l(s)$ 
by using equation~(\ref{eq:mul}) with the distant-observer approximation
from those obtained by considering the full redshift-space correlation
function in equation~(\ref{eq:fulld}). For simplicity, we assume that 
the bias factor of the galaxy samples is unity and constant in time. 
In computing $\MX_l(s)$ with the
distant-observer approximation, we consider three different cases: 
mean-redshift (dotted), volume-weighted (dashed), and pair-weighted (solid).
As the simplest case (dotted), we compute $\MX_l(s)$ by evaluating 
$\PZ(k,\mu_k)$ in equation~(\ref{eq:kaiser1})
at the mean redshift, accounting 
for the change in the logarithmic growth rate~$f$ and the growth factor 
in the matter power spectrum $\PM(k|\bar z)$. As the next case and
a better representation 
of the measurements, we average the correlation function $\MX_l(s)$ with the 
distant-observer approximation 
over the survey volume (dashed). As our last case with the highest 
sophistication (solid), we compute $\MX_l(s)$
by evaluating equation~(\ref{eq:fulld}), but with the full redshift-space
correlation function $\XZ(s,\phi_1,\phi_2)$
replaced by the simple Kaiser formula $\XZ(s,\mu)$ 
in equation~(\ref{eq:zcorr}). While the simple Kaiser formula is independent
of~$\oo$, we evaluate $\XZ(s,\mu)$ at the mean redshift of the pair, 
accounting for the change of pairs at different redshifts.

The three different ways (various curves)
of computing $\MX_l(s)$ with the distant-observer approximation yield nearly
identical results for the SDSS  samples in the top panels of 
Fig.~\ref{fig:powSDSS}.
The monopole $\MX_0(s)$ becomes negative around $s\sim130\hmpc$, while
the quadrupole $\MX_2(s)$ is negative and the hexadecapole $\MX_4$ 
is positive on all scales.
The differences in the multipole correlation functions 
are $\ll10^{-3}$ on all scales, which
is less than a percent level of $\MX_l(s)$ for $s\leq200\hmpc$.
However, on large scales $s\gg200\hmpc$ the correlation fucntion itself
is very small, such that the fractional deviation becomes large $>10\%$.
Gray curves show the calculation for the SDSS LRG sample
with the FKP weighting. No substantial difference is present.
With the FKP weighting,  the mean redshift is reduced to $\bar z=0.27$
from $\bar z=0.34$,
and the deviation from the simple Kaiser formula is slightly larger
than for the SDSS uniform sample.
Since the redshift range of the SDSS is rather narrow $z=0.15-0.45$,
three different ways of computing the simple Kaiser formula with the 
distant-observer approximation agree well with each other.

Bottom panels of Figs.~\ref{fig:powSDSS}$-$\ref{fig:powBIGB}
quantify the systematic errors in the monopole $\MP_0(k)$, the quadrupole
$\MP_2(k)$, and the hexadecapole $\MP_4(k)$ power spectra in the simple
Kaiser formula with the distant-observer approximation, where the
simple Kaiser formula is again computed by using three different methods
(various curves). 
Using equation~(\ref{eq:mulp}), the full redshift-space multipole
power spectra are computed by Fourier transforming the full redshift-space
multipole correlation functions,
and the error bars are obtained by using the covariance matrix in
equation~(\ref{eq:cov}). 

\begin{figure}
\centerline{\psfig{file=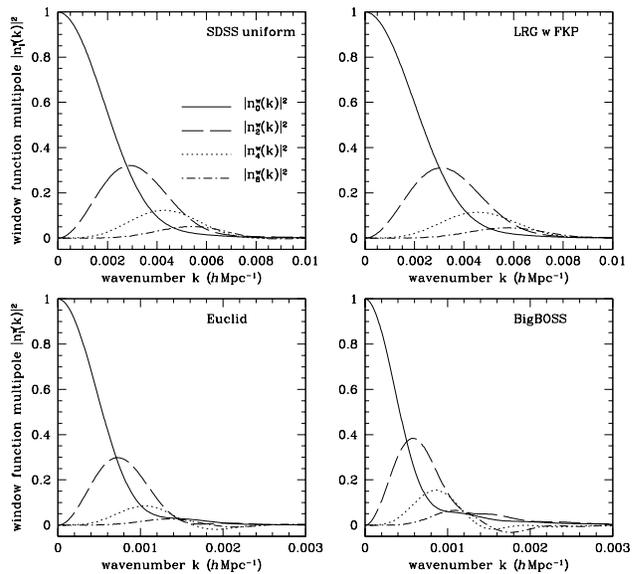, width=3.4in}}
\caption{Survey window function multipoles. The survey window function
$|n_g^w(\kvec)|^2$ is decomposed in terms of Legendre polynomials, and only
multipoles up to $l=6$ are plotted. For illustration, the window function 
is re-normalized to unity at $k=0$ for the monopole. While the top two panels
represent the same survey geometry, their window functions are slightly
different due to different weighting scheme.}
\label{fig:window}
\end{figure}

The deviation from the distant-observer approximation is again negligible 
for the SDSS samples on almost all scales (see, also, \citealt{SAPERA12}). 
It deviates at most four percents for the monopole
and eight percents for the quadrupole at the smallest wavenumber available 
in the survey, while the deviation in the hexadecapole is somewhat larger
than in the monopole and the quadrupole due to its sensitivity to larger
separation, $s\sim l/k$. 

Figures~\ref{fig:powEUCL} and~\ref{fig:powBIGB} show the systematic errors
of the simple Kaiser formula with the distant-observer approximation
in Euclid and the BigBOSS. Since galaxies in these surveys are farther away
from the observer than in the SDSS, the deviation in the multipole
correlation functions at a given separation is even smaller, 
despite their somewhat larger sky coverage than in the SDSS. In
Figs.~\ref{fig:powEUCL} and~\ref{fig:powBIGB} we now plot the fractional
deviations for the multipole correlation functions.
With larger redshift ranges in Euclid and the BigBOSS, three different ways
of computing the simple Kaiser formula start to diverge from each other. The
mean redshifts of the surveys are $\bar z=1.4$ and~2.8, and the redshift-space
multipole correlation functions at the mean redshift differ from those
averaged over the survey volume, because the former evaluates 
$b^2(\bar z)D^2(\bar z)$, $b(\bar z)f(\bar z)D^2(\bar z)$, 
$f^2(\bar z)D^2(\bar z)$
in equations~(\ref{eq:coef1})$-$(\ref{eq:coef3}) at the mean redshift, 
while the latter averages $b^2D^2$, $bfD^2$, $f^2D^2$ over the survey volume,
though the difference is less than 10\%.

The simplest method (dotted) of computing the multipoles at the mean redshift
yields the multipole correlation functions off in amplitude 
at the 2$-$4 percent level on small scales, but the shape is still consistent
with the full redshift-space correlation function. The volume-average
correlation functions (dashed) are better estimates of the full 
redshift-space correlation function, practically on all scales considered
in the plot. While the pair-weighted correlation functions (solid) perform
better than the volume-average for higher multipoles, the difference is rather
small, and it is computationally more expensive, as it needs to be computed
for each pair of galaxies.
The deviation in the multipole power spectra in the bottom panels are 
correspondingly small as in the multipole correlation functions in the
upper panels. 
Note again that the scales of the plot in the bottom
panels are different from those in the upper panels.
With more volume at higher redshifts, 
these surveys can probe larger scales (smaller $\kmin$), 
and the measurement uncertainties at a
fixed scale are substantially reduced by more than a factor of five.
However, as is evident in Figs.~\ref{fig:powEUCL} and~{\ref{fig:powBIGB},
the simple Kaiser formula with the distant-observer approximation
is in all practical purposes accurate in the future surveys.

\subsection{Wide angle effect in power spectrum analysis}
\label{ssec:pow}

We have discussed the wide angle effect in the correlation function and
the power spectrum measurements in the galaxy surveys. In our comparison,
the simple Kaiser formula with the distant-observer approximation is compared
to the full redshift-space correlation function in 
equation~(\ref{eq:fullcorr}). However,
since no analytic formula accounting for the velocity contribution and 
the wide angle effect is available for the redshift-space power spectrum,
the deviation in the multiple correlation functions 
from the simple Kaiser formula
is Fourier transformed to compute the deviation in the redshift multipole
power spectra. While this procedure should yield the correct result in 
an ideal case as the correlation function and the power spectrum are
Fourier counterparts, the power spectrum analysis in practice
is performed in a different way, and these two statistics contain 
not completely overlapping information as 
considered only over limited and different ranges of scale.

In general the power spectrum analysis is performed by using a weighted
quadratic function of the observed galaxy number density field,
and various methods in the power spectrum analysis differ in 
how they choose the pair-weighting function (see \citealt{TEHAET98}).
The standard method including the FKP method is the simplest, in which
the weighted number density field $n_g^w(\svec)=w(\svec)n_g(\svec)$
is Fourier transformed, and their amplitude is an estimate of the 
redshift-space power spectrum
\beeq
\hat\PZ(\kvec)\equiv|n_g^w(\kvec)|^2-P_\up{shot}(\kvec)~,
\label{eq:powest}
\eneq
where $P_\up{shot}(\kvec)$ represents the shot-noise contribution.
Again, we assume that the shot-noise contribution is negligible.

\begin{figure*}
\centerline{\psfig{file=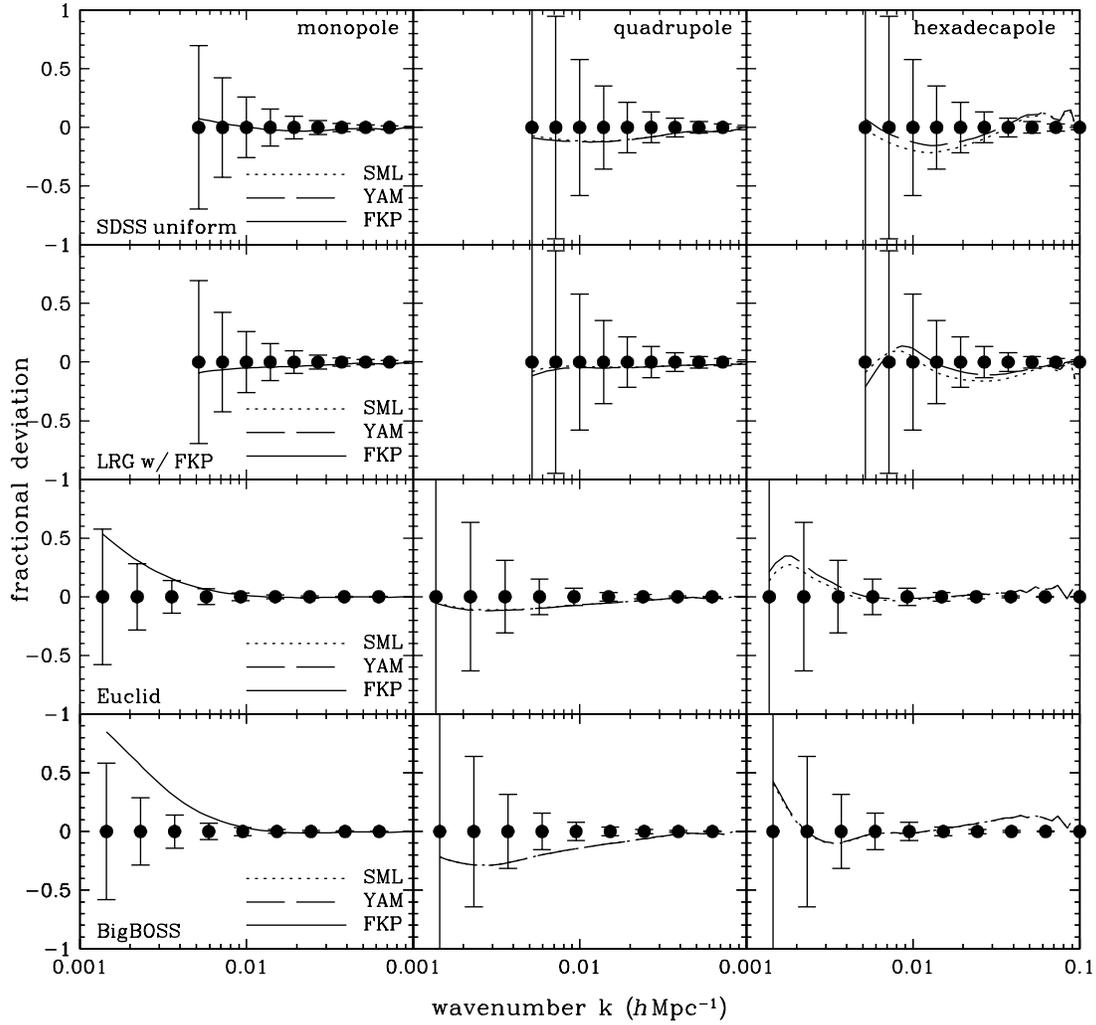, width=6.0in}}
\caption{Fractional deviation of estimates of the redshift-space multipole 
power spectra, relative to the simple Kaiser formula with the 
distant-observer approximation. All the redshift-space multipole power
spectra are scaled by a constant factor to match the monopole on small
scales for comparison. Each column shows the monopole, the quadrupole,
and the hexadecapoles, and each rows represents four different surveys.
The multipole power spectra with the
distant-observer approximation are volume-averaged over the survey volume
and convolved with the survey window function.
Three different curves represent different estimates of 
the redshift-space multipole power spectra. While these estimators
differ only in defining the line-of-sight direction for galaxy pairs
(FKP; \citealt{FEKAPE94}, YAM; \citealt{YANAET06}, SML; \citealt{SZMALA98}),
they are identical for the monopole power spectrum, and
two methods (YAM \& SML) yield similar results for higher multipoles, as their 
line-of-sight directions are nearly identical due to large distances to
galaxy pairs in surveys.}
\label{fig:powmea}
\end{figure*}

The FKP method is designed to measure the monopole power spectrum, {\it not}
the anisotropic redshift-space power spectrum.\footnote{An 
alternative to this trend
has been well developed \citep{BIQU91,FISCLA94,FILAET95,HETA95,FILAET95,
BAHETA95,HACU96},
in which the observed galaxy number density is decomposed in terms of radial
and angular eigenfunctions of the Helmholtz equation. The analysis is performed
on a sphere, explicitly accounting for each line-of-sight direction of 
galaxies. It forms the most natural basis for all-sky analysis, and
it becomes identical to the traditional 
power spectrum analysis on a small patchy of sky.
We refer the reader to the pedagogical review for all-sky
likelihood analysis \citep{HAMIL05}.
It has been applied to
various galaxy surveys in the past (\citealt{FILAET95,TABAET99,HATEPA00,
TABAET01,PATEHA01,TEHAXU02,PEBUET04,TESTET04,
TEEIET06,SHCRPE12}, see also \citealt{RARE12,PRMU13,YODE13}
for recent theoretical development). While the FKP method has been 
extensively used in recent galaxy surveys (e.g., \citealt{REPEET10,PEREET10}),
the spherical Fourier analysis provides an alternative, and this method is 
devoid of any wide angle effect (hence it is {\it not} the subject of current
investigation). However, the downside of this all-sky analysis is the 
ambiguity in relating the angular multipole~$l$ 
to the cosine angle~$\mu_k$ relative to the line-of-sight direction. 
On small scales, this ambiguity in the correspondence is
disadvantageous, as most theoretical 
models of nonlinear redshift-space distortion build on the distant-observer
approximation.}
Few exceptions are \cite{COFIWE94} and \cite{YANAET06} analysis of the 2dFGRS.
In the former, only pairs in the same but small angular patchy 
are correlated to 
ensure that those pairs share the same line-of-sight direction. With this
prescription, the power spectrum analysis, however, suffers from non-trivial
window function, aliasing of modes, and larger computing cost. No further
development or application to observation has been made along this direction.
In the latter, a simple method is proposed to account for the line-of-sight
direction change in the power spectrum analysis \citep{YANAET06},
while keeping the simplicity in the FKP method.
We adopt it to compare the 
difference among various methods in the power spectrum analysis.

In the standard FKP method,
the ensemble average of the power spectrum estimate in 
equation~(\ref{eq:powest}) is
\beeq
\left\langle\hat\PZ(\kvec)\right\rangle
=\int d^3\svec_1\int d^3\svec~e^{-i\kvec\cdot\svec}~
\bar n_g^w(\svec_1)~\bar n_g^w(\svec_2)~\XZ(\svec_1,\svec_2) ~,
\label{eq:fkppow}
\eneq
where the separation vector is $\svec=\svec_1-\svec_2$ and the full
redshift-space correlation function $\XZ(\svec_1,\svec_2)$ 
depends on the triangular configuration of the pair, including the opening
angle~$\oo$ seen from the observer.
Since we use the weighted number density $n_g^w(\svec)$
rather than the fluctuation~$\delta_g$, the normalization of the weight
function must be chosen to ensure that the power spectrum estimate is unbiased
\citep{FEKAPE94},
\beeq
\left\langle\hat\PZ(\kvec)\right\rangle=\int{d^3\kvec'\over(2\pi)^3}~
\PZ(\kvec')~|\bar n_g^w(\kvec-\kvec')|^2 ~,
\label{eq:pconv}
\eneq
and the normalization condition is
\beeq
1=\int{d^3\kvec\over(2\pi)^3}|\bar n_g^w(\kvec)|^2
=\int d^3\svec\left[\bar n_g^w(\svec)\right]^2~.
\eneq
The weighted mean number density $\bar n_g^w$ is often referred to as the
survey window function, and its dimension is $[\bar n_g^w(\kvec)]=V^{1/2}$
and $[\bar n_g^w(\svec)]=V^{-1/2}$.
Figure~\ref{fig:window} delineates the survey window functions. The window
functions of four different surveys are largely self-similar, as each survey
differs only in the redshift depth and sky coverage.
While the survey window function is anisotropic, it can
be conveniently decomposed in terms of its multipole functions due to the
azimuthal symmetry of the survey geometries we considered in the paper.
The monopole (solid) reflects the characteristic scale set by the survey 
volume, and higher multipoles account for the anisotropy of the survey 
geometry. When the survey volume is infinite, the monopole becomes the
Dirac delta function, and the higher multipoles vanish.

Using equation~(\ref{eq:multi}),
the multipole power spectra are obtained by angle-averaging the redshift-space
power spectrum in equation~(\ref{eq:fkppow}) as
\bear
\label{eq:mulfkp}
\left\langle \hat\MP_l(k)\right\rangle&=&(-i)^l(2l+1)\int d^3\svec_1
\int d\ln s~s^3j_l(ks)\\
&\times&\int d^2\hat\svec~\LL_l(\hat\zvec\cdot\hat\svec)~
\bar n_g^w(\svec_1)~\bar n_g^w(\svec_2)~\XZ(\svec_1,\svec_1-\svec)~, \nonumber
\enar
where the line-of-sight direction is set to be $\hat\zvec$-direction
and we used
\beeq
\int d^2\hat\kvec~e^{-i\kvec\cdot\svec}~\LL_l(\Vang\cdot\hat\kvec)
=4\pi(-i)^l~j_l(ks)~\LL_l(\Vang\cdot\hat\svec)~.
\eneq
With the distant-observer approximation, the angular integral 
in equation~(\ref{eq:mulfkp}) becomes
\beeq
\int d^2\hat\svec~\LL_l(\hat\zvec\cdot\hat\svec)~\XZ(\svec_1,\svec_1-\svec)
\RA{4\pi\over2l+1}~\MX_l(s)~,
\eneq
and we recover equation~(\ref{eq:mulp}) for the multipole power spectra
$\MP_l(k)$. It becomes clear in equation~(\ref{eq:mulfkp})
that the standard FKP method cannot be used to measure the anisotropic
redshift-space power spectrum because the line-of-sight
direction is fixed as $\hat\zvec$-direction, independent of the pair
configuration. This is the reason that the FKP method is used solely for
the monopole power spectrum, as is originally intended. However, one issue
is that the redshift-space correlation function is Fourier
transformed without accounting for the non-uniform distribution of the
cosine angle~$\mu$.

It is apparent in equation~(\ref{eq:mulfkp}) that the redshift-space 
multipole power
spectrum is estimated without accounting for the pair distribution.
While the one-point window functions $\bar n_g^w(\svec_i)$ are considered, 
the window function for the pair distribution (or the two-point window 
function)  is not considered in estimating the redshift-space power spectrum. 
For example, the monopole 
power spectrum will be estimated by summing all pairs within the survey 
weighted by the window function, such that it will be biased if the geometry
is like a pencil beam, because pairs within the survey are mostly
separated along the line-of-sight direction. Naturally, this issue 
gives rise to negligible systematic errors on small separations compared to
the survey scale, but we show in Figure~\ref{fig:powmea} that 
it can cause significant systematic errors on large scales.

Due to the simplicity of the FKP method, 
\cite{YANAET06} attempt to extend the FKP method for the anisotropic
redshift-space power spectrum by using a pair-dependent direction
$\svec_h\equiv(\svec_1+\svec_2)/2$ as the 
line-of-sight direction for each pair, such that the estimators for the
redshift-space multipole power spectra are
\beeq
\hat P^\up{Y}_l(\kvec)\equiv
\int d^3\svec_1\int d^3\svec~e^{-i\kvec\cdot\svec}~
n_g^w(\svec_1)~n_g^w(\svec_2)~\LL_l(\hat\svec_h\cdot\Kang)~.
\eneq
Their ensemble averages are
\bear
\label{eq:mulyam}
\left\langle \hat P^\up{Y}_l(k)\right\rangle&=&(-i)^l(2l+1)\int d^3\svec_1
\int d\ln s~s^3j_l(ks)\\
&\times&\int d^2\hat\svec~\LL_l(\hat\svec_h\cdot\hat\svec)
\bar n_g^w(\svec_1)~\bar n_g^w(\svec_2)~\XZ(\svec_1,\svec_1-\svec)~, \nonumber
\enar
and we recover the same limit with the distant-observer approximation.
As opposed to the standard
FKP method, the Yamamoto method is 
computationally more expensive, because one cannot simply Fourier transform
the weighted number density~$n_g^w$ and compute its amplitude
to estimate the redshift-space multipole
power spectrum, instead it has to be estimated by a pair-dependent way.
As another way of measuring the redshift-space multipole power spectra,
we also consider a variant of the Yamamoto method by using
the line-of-sight direction~$\Vang$ that bisects the pair in angle (see
Fig.~\ref{fig:geometry}). All of the three FKP-based
methods agree for the 
monopole power spectrum, as the angle average of the wavevector 
removes the need to define the line-of-sight direction, and
as is intended in the original FKP method.
Regarding the non-uniform distribution of cosine angle~$\mu$,
this issue is not addressed
in any of the three methods for measuring the redshift-space 
multipole power spectra, and we find that substantial systematic errors arise
from this issue in future surveys.

Figure~\ref{fig:powmea} compares the three different estimates of the
redshift-space multipole power spectra to the simple Kaiser formula with
the distant-observer approximation. 
We compute the redshift-space multipole power spectra for each method
by generating random catalogues of galaxies and by considering a spherical
volume integral of the full redshift-space correlation function 
$\XZ(\svec_1,\svec_2)$ in equations~(\ref{eq:mulfkp}) and~(\ref{eq:mulyam})
from each galaxy in the catalogues. For comparison, we obtain the 
theoretical prediction by convolving the simple Kaiser formula 
at the mean redshift with the survey window function as in 
equation~(\ref{eq:pconv}). Four different surveys are displayed in 
four rows, and the redshift-space multipole power spectra are presented in
three columns. 

In the previous section, the redshift-space correlation function is
averaged over the different opening angle~$\oo$, the multipole correlation
functions $\MX_l(s)$
are then obtained by accounting for the non-uniform distribution
of cosine angle~$\mu$, and they are finally Fourier transformed
to obtain the redshift-space multipole power spectra $\MP_l(k)$. However, in 
equation~(\ref{eq:fkppow}), the power spectrum receives contributions
with equal weight from all the correlation function at a given $(s,\mu)$ 
but with different opening angle~$\oo$, which will change the normalization
even on smallest scales, where the distant-observer approximation is accurate.
Therefore, we scale the monopole power spectrum estimate
with a constant factor to match the convolved monopole power spectrum
with the distant-observer approximation at $k=0.1\hmpci$, but 
the same scale factor is used to scale the quadrupole and the hexadecapole
power spectra. These scale factors result in $10-20\%$ shifts in amplitude
for each survey, but they can be absorbed into unknown galaxy bias factor.

The left column shows the monopole power spectrum for each survey, and the
three different methods yield identical results for the monopole.
The deviation of the monopole estimates in the SDSS is negligible, compared
to the sample variance, and the simple Kaiser formula with the distant-observer
approximation provides a good approximation to the monopole power spectrum
measurement. As galaxies in Euclid and the BigBOSS are farther away than in
the SDSS, the wide angle effect becomes irrelevant in these surveys. However,
while the survey volume in Euclid and the BigBOSS is dramatically larger 
than in the SDSS, the sky coverage increases only a little, especially so
in the BigBOSS. This affects the distribution of cosine angle~$\mu$ at the
largest separations, and the deviation in the monopole power spectrum is
larger than the sample variance, a potential systematic error that needs to
be addressed in the future surveys.

The middle and the right columns show the quadrupole and the hexadecapole
power spectra. Note that the standard FKP method is not intended to
measure the anisotropic redshift-space power 
spectrum.\footnote{\cite{BLBRET11b} 
measured the WiggleZ 
redshift-space power spectrum by using both the standard
FKP and the 
Yamamoto methods. Although no significant detection is made for the 
hexadecapole power spectrum $P_4(k)$ in the WiggleZ survey, they find that 
both methods
yield measurements of the monopole and the quadrupole power spectra, consistent
with each other. While the sky coverage in the WiggleZ survey is
$\sim1000$~deg$^2$, it is divided into six disjoint regions, each of which
is subtended by $\sim10$~deg. Therefore, as noted in \cite{BLBRET11b},
a single line-of-sight 
assignment to all galaxies in each region is a good approximation in the
WiggleZ survey.}
Two methods correctly define the line-of-sight direction
for each galaxy pair and yield virtually identical results for the quadrupole
and the hexadecapole. As opposed to the monopole power spectrum, 
where the deviations accumulate, the 
effect of the non-uniform distribution of~$\mu$ is smaller for the quadrupole
and the hexadecapole, as the higher order Legendre polynomials
oscillate around zero. There exist small but wiggly deviations on small scales
for the hexadecapole power spectrum. On those scales, the distant-observer
approximation is accurate and the $\mu$-distribution is uniform. We suspect
that the few percent level deviations on small scales 
are numerical artifact, arising from 
the complication of multi-dimensional integration in 
equation~(\ref{eq:pconv}) and the sensitivity to angular structure in the
survey window function.

\section{DISCUSSION}
\label{sec:discussion}
We have investigated the wide angle effects in galaxy clustering 
measurements in galaxy surveys such as the SDSS, Euclid and 
the BigBOSS. We have found that compared to the measurement uncertainties
associated with the redshift-space multipole correlation functions
the wide angle effects are negligible in the SDSS 
(as discussed in \citealt{SAPERA12})
and they are completely irrelevant in the future surveys, provided that
the Kaiser formula is averaged over the survey volume to represent the
redshift-space correlation function at the representative redshift.
While we have reached the same conclusion in the power spectrum analysis,
the standard power spectrum analysis based on the FKP method
is performed in a slightly different
way, carrying a systematic flaw that can 
result in substantial systematic errors in the future surveys. Furthermore,
we have clarified a connection of the Kaiser formula to the relativistic
formula of galaxy clustering, and have found corrections to the formula
often used for computing the correlation function of widely separated pairs.

In literature,
the ``wide angle effects'' are often used to refer to the deviation 
from the simple Kaiser formula with the distant-observer approximation, 
but they arise owing to two physically distinct effects. First, the two-point
correlation function involves a triangle formed by a galaxy pair and the 
observer, and the triangle is described by two line-of-sight directions and
the pair separation. The two line-of-sight directions of each galaxy
deviate from the one and only line-of-sight direction in the
distant-observer approximation, and the resulting correlation function 
depends on the triangular configuration, whereas the simple Kaiser formula
is just a function of the pair separation and the angle it makes with the
line-of-sight direction. This effect is legitimately called the wide angle
effect. 
The other effect that contributes to the deviation from the simple Kaiser
formula is the velocity contribution to galaxy clustering in redshift-space. 
Even in linear theory, the velocity terms are present in the full Kaiser
formula, arising from the mapping of galaxy number densities
in real-space to redshift-space. Moreover, the relativistic treatment
of galaxy clustering reveals the presence of the gravitational
potential contribution to galaxy clustering, though their corrections are
even smaller than the velocity contribution. This effect is often referred
to as the wide angle effect, but is independent of how widely galaxy pairs 
are separated on the sky.

We have made a connection of the full Kaiser formula to the 
relativistic description of galaxy clustering
\citep{YOHAET12}. The full Kaiser formula is valid up to the velocity 
contribution, only when galaxy samples are selected without any bias,
except one from the observed redshift.  
Furthermore, we have found that even the full Kaiser formula
is not properly considered in the formula for computing
the redshift-space correlation function \citep{SZMALA98,SZAPU04,PASZ08}.
In light of the relativistic treatment, the derivative in the Jacobian 
mapping to redshift-space involves the time derivative as well as the
spatial derivative, as we only observe galaxies in the past light cone.
The (missing) time derivative give corrections to the velocity
contribution. While these terms again contribute to negligible systematic 
errors, they can be readily implemented to the existing
formula by modifying the selection function~$\alpha$ (see, Sec.~\ref{ssec:gr}
for details).

The main reason that the deviation from the distant-observer approximation
is small is already evident in Fig.~\ref{fig:wide}: Galaxies in typical
surveys are substantially far away from the observer, and there
are simply {\it not} many galaxies in the local neighbourhood, where
the opening angles of galaxy pairs can be large. For example, typical
galaxy pairs at a separation $s\simeq100\hmpc$ 
in the SDSS would appear subtended by less than 5~degrees on the sky,
and this trend is accelerated in future surveys at higher redshifts,
where typical distances to galaxy samples in those surveys are at least
twice larger than in the SDSS.
A similar argument can be made to attempt to reach the opposite conclusion:
Given a typical distance~$r$ to galaxy pairs in a survey, the deviation
from the distant-observer approximation becomes substantial as widely
separated galaxy pairs are considered. This argument is in fact true,
and non-negligible deviations ($>10\%$) in the multipole correlation 
functions in Fig.~\ref{fig:powSDSS} exist at $s>400\hmpc$, while the
deviations in future surveys at higher redshifts are substantially smaller
(see, Figs.~\ref{fig:powEUCL} and~\ref{fig:powBIGB}). However, the
deviations are smaller than one naively expects: At a given pair
separation~$s$ and cosine angle~$\mu$, the correlation
function $\XZ(s,\mu)$ is obtained by averaging the full
correlation function $\XZ(s,\mu,\oo)$ over all galaxy pairs with different
opening angles~$\oo$. With more volume at a larger distance from the observer,
the correlation function is skewed to that of galaxy pairs farther away from
the observer, of which the opening angles are smaller. 

However small, the deviations may matter in the era of precision cosmology.
To quantify the systematic errors associated with the distant-observer
approximation, we have computed the covariance matrix of the redshift-space
multipole power spectra, obtained by Fourier transforming the 
redshift-space multipole correlation functions, accounting for the deviation
from the distant-observer approximation. 
In agreement with \cite{SAPERA12}, we have found that the systematic
errors by using the distant-observer approximation are
negligible in the SDSS on all scales (see Fig.~\ref{fig:powSDSS}), provided
that the Kaiser formula with the distant-observer approximation
is properly evaluated. It is found that averaging the prediction of
the Kaiser formula over the survey volume yields the best match to the
full redshift-space clustering measurements, while the evaluation at the mean
redshift of the survey would be just fine at the few percent level in 
amplitude (the shape is correctly estimated in both cases).
Furthermore, this conclusion remains valid in future surveys such as Euclid and
the BigBOSS at higher redshifts
(Figs.~\ref{fig:powEUCL} and~\ref{fig:powBIGB}). At a given pair 
separation (with fixed sky coverage), 
one would need a survey at a higher redshift to reduce
measurement uncertainties associated with that scale by sampling more
independent Fourier modes, 
$\sigma\propto1/\sqrt{N_k}\propto1/\sqrt{V}\propto1/r$. 
However, since the deviation from the
distant-observer approximation is to the least, $\sin\oo\simeq\oo\propto1/r$
or $|\cos\oo-1|\simeq\oo^2\propto1/r^2$, the systematic errors in the 
distant-observer approximation can only decrease in a survey at 
higher redshifts, and they are already negligible in the SDSS.

While the distant-observer approximation is accurate in those surveys,
the power spectrum measurements have an issue in the
standard power spectrum analysis. Although 
negligible in the SDSS, we have found 
that the non-uniform distribution of cosine 
angle~$\mu$ between the line-of-sight and the pair separation directions
is {\it not} properly considered in the standard  power spectrum analysis and
can be a substantial source of systematic errors in the future surveys
(see Fig.~\ref{fig:powmea}).

Our conclusion should be taken with caution that systematic errors
incurred by adopting the distant-observer approximation for computing
the redshift-space galaxy clustering are negligible 
in generic galaxy redshift surveys such as the SDSS, Euclid, and the BigBOSS,
provided that the correlation function or the power spectrum is obtained
by properly averaging over all galaxy pairs in the surveys. 
Certainly, one could
divide survey regions into multiple redshift bins with narrow width to
prevent further dilution of the correlation function or the power spectrum
due to averaging over galaxy pairs at farther
distances (see, e.g., \citealt{MODU12,ASCRET12}). However, it still remains
to be demonstrated whether the systematic errors of the distant-observer
approximation in those survey configurations can be important, because
smaller volumes in those redshift bins lead to larger measurement 
uncertainties.

Alternatively, instead of using the currently popular methods discussed 
in this paper, one can choose to use the maximum likelihood methods and
measure the correlation function
$\XZ(s,\phi_1,\phi_2)$ or the spherical power spectrum based on spherical
Fourier decomposition (e.g., \citealt{FISCLA94,HETA95,TEBLET04,HAMIL05}).
Under the approximation that the underlying distribution is Gaussian, 
this will circumvent the issues of the wide angle effect and the non-uniform 
distribution on large scales. Further decomposition into the standard
redshift-space multipole correlation function $\MX_l(s)$ or power spectrum 
$\MP_l(k)$ would require the assumption
that the observed correlation function or spherical power spectrum be well
described by the simple Kaiser formula with the distant-observer approximation,
which we showed is a good approximation. We suspect that maximum likelihood
methods may provide optimal ways to measure large-scale clustering without
significant issues discussed in this paper.


\section*{acknowledgments}
We acknowledge useful discussions with Florian Beutler, Eyal Kazin, 
Shun Saito, Anze Slosar, and Martin White. 
This work is supported by the Swiss National Foundation (SNF) under contract
200021-116696/1 and WCU grant R32-10130.
J.Y. is supported by the SNF Ambizione Grant.

\bibliography{../../reference}

\appendix   
\section{Covariance matrix of multipole power spectra}
\label{app:cov}
With the distant-observer approximation, the redshift-space power spectrum
is well described by the simple 
Kaiser formula in equation~(\ref{eq:kaiser1}) in 
the linear regime, and it consists of only three multipole power spectra. 
While the covariance matrix of the multipole power spectra is diagonal in 
Fourier modes $\propto\delta_{kk'}$, they are correlated between different 
angular multipoles. Using equation~(\ref{eq:covariance}), 
the diagonal components of the covariance matrix can be computed as
\bear
\up{Cov}[\hat\MP_0\hat\MP_0]&=&[\TMP]^2+{1\over5}
[\MP_2]^2+{1\over9}[\MP_4]^2\\
&=&{1\over\bng^2}+{2\over\bng}\left(1+{2\beta\over3}+{\beta^2\over5}\right)
b^2\PM\nonumber\\
&&+\left(1+{4\beta\over3}+{6\beta^2\over5}+{4\beta^3\over7}
+{\beta^4\over9}\right)[b^2\PM]^2~, \nonumber \\
\up{Cov}[\hat\MP_2\hat\MP_2]&=&5[\TMP]^2+{20\over7}\TMP\MP_2
+{15\over7}[\MP_2]^2+{20\over7}\TMP\MP_4 \nonumber \\
&&+{120\over77}\MP_2\MP_4+{8945\over9009}[\MP_4]^2 \\
&=&{5\over\bng^2}+{10\over\bng}\left(1+{22\beta\over21}+{3\beta^2\over7}
\right)b^2\PM\nonumber\\
&+&5\left(1+{44\beta\over21}+{18\beta^2\over7}+{340\beta^3\over231}
+{415\beta^4\over1287}\right)[b^2\PM]^2~,\nonumber \\
\up{Cov}[\hat\MP_4\hat\MP_4]&=&9[{\tilde \MP_0}]^2+{360\over77}\TMP\MP_2
+{16101\over5005}[\MP_2]^2 \\
&&+{2916\over1001}\TMP\MP_4
+{3240\over1001}\MP_2\MP_4+{42849\over17017}[\MP_4]^2 \nonumber\\
&=&{9\over\bng^2}+{18\over\bng}\left(1+{78\beta\over77}+{1929\beta^2\over5005}
\right)b^2\PM\nonumber\\
&&\hspace{-30pt}
+9\left(1+{156\beta\over77}+{11574\beta^2\over5005}+{1308\beta^3\over1001}
+{711\beta^4\over2431}\right)[b^2\PM]^2~,\nonumber \\
\enar
where $\TMP\equiv\MP_0+1/\bng$ and
we suppressed the scale-dependence of the multipole power spectra and
their covariance matrix. In \cite{TANISA10}, there were minor typos 
in their equation~(C4) for $\up{Cov}[\hat\MP_2\hat\MP_2]$.
The off-diagonal components of the covariance matrix are
\bear
\up{Cov}[\hat\MP_0\hat\MP_2]&=&2\TMP\MP_2+{2\over7}[\MP_2]^2
+{4\over7}\MP_2\MP_4+{100\over693}[\MP_4]^2  \\
&=&{8\beta\over\bng}\left({1\over3}+{\beta\over7}\right)b^2\PM\nonumber\\
&&+8\beta\left({1\over3}+{3\beta\over7}+{5\beta^2\over21}+{5\beta^3\over99}
\right)[b^2\PM]^2~,\nonumber \\
\up{Cov}[\hat\MP_0\hat\MP_4]&=&{18\over35}[\MP_2]^2+2\TMP\MP_4+{40\over77}\MP_2
\MP_4+{162\over1001}[\MP_4]^2  \\
&=&{16\beta^2\over35~\bng}b^2\PM+48\beta^2\left({1\over35}+{2\beta\over77}
+{\beta^2\over143}\right)[b^2\PM]^2~,
\nonumber \\
\up{Cov}[\hat\MP_2\hat\MP_4]&=&{36\over7}\TMP\MP_2+{108\over77}[\MP_2]^2
+{200\over77}\TMP\MP_4 \nonumber \\
&&+{3578\over1001}\MP_2\MP_4+{900\over1001}[\MP_4]^2  \\
&=&\left({48\beta\over7~\bng}+{272\beta^2\over77\bng}\right)b^2\PM\nonumber\\
&&+\left({48\beta\over7}+{816\beta^2\over77}+{6960\beta^3\over1001}
+{240\beta^4\over143}\right)[b^2\PM]^2~.\nonumber
\enar

\begin{figure}
\centerline{\psfig{file=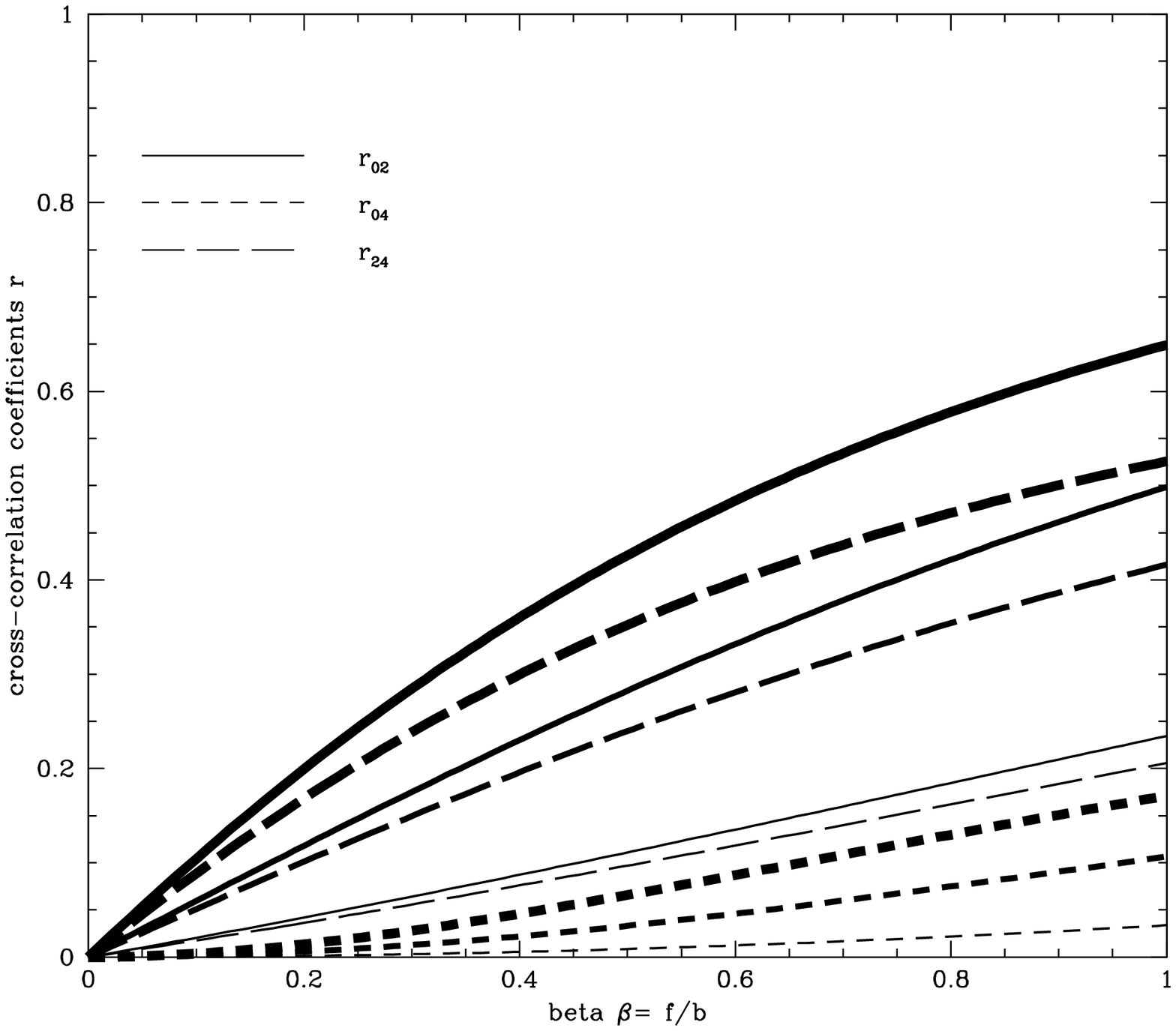, width=3.4in}}
\caption{Cross-correlation of the multipole covariance matrices. As a function
of~$\beta=f/b$, various curves show the cross-correlation coefficients
$r_{02}$ (solid), $r_{04}$ (dashed), $r_{24}$ (short dashed). With increasing
thickness, each curve represents galaxy samples with $\bng P_g=0.2$
(thin), 1~(thicker), and~10 (thickest).}
\label{fig:cross}
\end{figure}

The degree to which the covariance matrix of the multipole power spectra is
correlated is described by the cross-correlation coefficients:
\beeq
r_{l_1l_2}={\up{Cov}[\hat\MP_{l_1}\hat\MP_{l_2}]\over\sqrt{\up{Cov}
[\hat\MP_{l_1}\hat\MP_{l_1}]~\up{Cov}[\hat\MP_{l_2}\hat\MP_{l_2}]}}~,
\eneq
ranges from zero (uncorrelated) to unity (perfectly correlated).
Figure~\ref{fig:cross} shows the cross-correlation coefficients.
Three different curves show the coefficients $r_{02}$ (solid), $r_{04}$
(dashed), and $r_{24}$ (short dashed), respectively. For each curve, there
exist three different thickness, by which three ratios of the
sample variance to the shot noise are considered. Throughout the paper,
we considered the sample variance limited case $\bng P_g=\infty$,
in which the galaxy number density is high and the shot noise contribution 
is negligible. The cases with $\bng P_g=10$ (thickest), 1~(thicker),
and~0.2 (thin) approximately
correspond to the SDSS LRG sample at $k=0.01$, $0.15$, and $0.3~\hmpci$,
respectively. 
All of three cases show that the 
covariance matrix becomes uncorrelated in the limit of highly biased
galaxy sample ($\beta\RA0$).

\end{document}